\documentclass[fleqn,usenatbib,usedcolumn]{mnras}

\usepackage{newtxtext,newtxmath}

\usepackage[T1]{fontenc}
\usepackage{ae,aecompl}

\usepackage{multirow}
\usepackage{caption}
\usepackage{dsfont}
\usepackage[capitalise]{cleveref}
\usepackage{color}
\usepackage{arydshln}
\usepackage{commath}

\usepackage{graphicx}

\usepackage{etoolbox}
\makeatletter
\patchcmd\@combinedblfloats{\box\@outputbox}{\unvbox\@outputbox}{}{%
  \errmessage{\noexpand\@combinedblfloats could not be patched}%
}%
\makeatother

\usepackage{graphicx}	
\usepackage{amsmath}	
\usepackage{amssymb}	
\usepackage{csquotes}

\newcommand{\Gyr}{\,\hbox{Gyr}}

\newcommand{\Msol}{\,{\hbox{M}_\odot}}
\newcommand{\noop}[1]{}
\newcommand{\ro}[1]{\ensuremath{\textrm{#1}}}

\newcommand{\dd}{\ensuremath{\ro{d}}}

\title[The impact of $\Lambda$ on galaxy formation]{The impact of dark energy on galaxy formation. What does the future of our Universe hold?}

\author[J. Salcido et al.]{Jaime Salcido,$^{1}$\thanks{E-mail:
\href{mailto:jaime.salcido@durham.ac.uk}{jaime.salcido@durham.ac.uk}} 
Richard G. Bower,$^{1}$ 
Luke A. Barnes,$^{2}$ 
Geraint F. Lewis,$^{2}$ \newauthor 
Pascal J. Elahi,$^{3}$  
Tom Theuns,$^{1}$ 
Matthieu Schaller,$^{1}$ 
Robert A. Crain,$^{4}$ \newauthor 
Joop Schaye$^{5}$ 
\\ \\
$^{1}$ Institute for Computational Cosmology, Department of Physics, Durham University, South Road, Durham, DH1 3LE, UK \\
$^2$ Sydney Institute for Astronomy, School of Physics, A28, The University of Sydney, NSW 2006, Australia \\
$^3$ Int Centre for Radio Astronomy Research, The University of Western Australia, 35 Stirling Highway, Crawley WA 6009, Australia\\
$^{4}$ Astrophysics Research Institute, Liverpool John Moores University, 146 Brownlow Hill, Liverpool L3 5RF, UK\\
$^{5}$ Leiden Observatory, Leiden University, P.O. Box 9513, 2300 RA Leiden, the Netherlands}

\date{Accepted XXX. Received YYY; in original form ZZZ}

\pubyear{2015}

\begin{document}
\label{firstpage}
\pagerange{\pageref{firstpage}--\pageref{lastpage}}
\maketitle

\begin{abstract}
We investigate the effect of the accelerated expansion of the Universe due to a cosmological constant, $\Lambda$, on the cosmic star formation rate. We utilise hydrodynamical simulations from the \textsc{Eagle} suite, comparing a $\Lambda$CDM Universe to an Einstein-de Sitter model with $\Lambda=0$. Despite the differences in the rate of growth of structure, we find that dark energy, at its observed value, has negligible impact on star formation in the Universe. We study these effects beyond the present day by allowing the simulations to run forward into the future ($t>13.8$ Gyr). We show that the impact of $\Lambda$ becomes significant only when the Universe has already produced most of its stellar mass, only decreasing the total co-moving density of stars ever formed by ${\approx}15\%$. We develop a simple analytic model for the cosmic star formation rate that captures the suppression due to a cosmological constant. The main reason for the similarity between the models is that feedback from accreting black holes dramatically reduces the cosmic star formation at late times. Interestingly, simulations without feedback from accreting black holes predict an upturn in the cosmic star formation rate for $t>15$ Gyr due to the rejuvenation of massive ($ > 10^{11} \mathrm{M}_{\sun}$) galaxies. We briefly discuss the implication of the weak dependence of the cosmic star formation on $\Lambda$ in the context of the anthropic principle.
\end{abstract}




\section{Introduction}\label{sec:intro}

Precise observational data from the past two decades has allowed us to measure the cosmic history of star formation back to very early times ($z \approx 8$). The star formation rate (SFR) density of the Universe peaked approximately 3.5 Gyr after the Big Bang ($z \approx 2$), and declined exponentially thereafter \citep[for a review see][]{Madau:2014}. 

Galaxy formation and evolution is a highly self-regulated process, in which galaxies tend to evolve towards a quasi-equilibrium state where the gas outflow rate balances the difference between the gas inflow rate and the rate at which gas is locked up in stars and black holes (BHs) \citep[e.g.][]{WF:1991,Finlator:2008, Schaye_et_al:2010, Dave:2012}. Consequently, the cosmic SFR density is thought to be determined both by the formation and growth of dark matter haloes, and by the regulation of the gas content in these haloes. The former depends solely on cosmology, whereas the latter depends on baryonic processes such as radiative cooling, stellar mass loss, and feedback from stars and accreting black holes.

Which of these factors is most responsible for the decline in cosmic star formation? It could be driven by the `freeze out' of the growth of large-scale structure, caused by the onset of accelerating cosmic expansion. As galaxies are driven away from each other by the repulsive force of dark energy, accretion and merging is slowed and galaxies are gradually starved of the raw fuel for star formation. Or, it could be caused primarily by the onset of efficient stellar and BH feedback. 

The discovery of the accelerating expansion of the Universe was a breakthrough achievement for modern cosmology \citep{SN2,SN1}. However, the driving force behind the acceleration (generically known as dark energy) is still unknown. At present, all cosmological observations are consistent with a cosmological constant, or a form of energy whose density remains constant as the Universe expands. One such form of energy is vacuum energy: the energy of a quantum field in its ground state (zero particles). The present best-fit cosmological model, known as the concordance model, or $\Lambda$CDM, includes both a cosmological constant $\Lambda$ and Cold (i.e. non-relativistic) Dark Matter. This model has been very successful in matching the observational data.

Nevertheless, the model raises a number of fundamental problems. Predictions from quantum field theory for the vacuum energy density overestimate the observed value of $\Lambda$ by many orders of magnitude \citep[for a review see][]{Weinberg:1989R,Carroll:2001}. In addition, the energy density of matter and the cosmological constant are within a factor of a few of each other at the present time, making our epoch unusual in the evolution of the Universe. This is known as the coincidence problem. These problems have motivated the search for alternative models of dark energy and modifications of gravity that might explain the acceleration of the universe more naturally. For example, quintessence models propose that the density of matter and dark energy track each other. In many models, however, fine tuning of the model parameters is still required to explain their observed similarity \cite[see for example][]{Weinberg:2000}.

An alternative approach is therefore to explain the observed value of $\Lambda$ on anthropic grounds. This has already been applied very promisingly to the coincidence problem. Since the coincidence concerns the time that \emph{we} observe the universe, the nature and evolution of observers in the Universe is highly relevant. For example \cite{Lineweaver:2007} argue that the production of planets in our Universe peaks when matter and dark energy are roughly coincident \citep[see, however,][]{Loeb:2016}.

For the cosmological constant and other fundamental parameters, anthropic reasoning requires a \emph{multiverse}. Many models of inflation, such as eternal inflation, imply that the Universe as a whole is composed of a vast number of inflationary patches or sub-universes. Each sub-universe inherits a somewhat random set of physical constants and cosmic parameters from a wide range of possible values. Sub-universes in which the cosmological constant is large and positive will expand so rapidly that gravitational structures, such as galaxies, are unable to form \citep[e.g.][]{Weinberg:1987Anthropic, Martel:1998, Efstathio:1995, Sudoh:2017}. Large negative values will cause the universe to collapse rapidly, also preventing the formation of galaxies. Only sufficiently small values of $\Lambda$ will lead to the formation of universes that are able to host observers. This argument eliminates extreme values of $\Lambda$. For example, \cite{Weinberg:2000} estimates an upper bound on a positive vacuum energy density to allow for the formation of galaxies of about 200 times the present mass density.

Refining Weinberg's estimate requires us to more accurately explore the sensitivity of galaxy formation to the presence of $\Lambda$. Here, we use a suite of hydrodynamical simulations to take a first look at this problem by calculating the effect of the cosmological constant on galaxy and star formation in our Universe. Specifically, we compare the formation of galaxies in our Universe with a hypothetical universe that is indistinguishable from ours at early times but has no cosmological constant. Because $\Lambda$ is negligibly small in the early universe, these two universes will evolve in nearly identical ways for the first $\approx 2$ Gyr of cosmic time (when the dark energy density is less than 0.03 times the matter density). This means that the epochs of nucleosynthesis, recombination\footnote{Of course, an observer in a $\Lambda=0$ universe would measure a different angular power spectrum in the cosmic microwave background after 13.8 Gyr, because of the very different expansion history of the Universe at later times.}, and reionization are indistinguishable.

In recent years, the accuracy of our understanding of galaxy formation has improved considerably, reaching the point at which it is possible to undertake this comparison meaningfully. The increased realism of simulated galaxies (in particular disc galaxies with more realistic sizes and masses) has been achieved due to the use of physically motivated subgrid models for feedback processes \citep[e.g.][]{Schaye:2015,Dubois:2016,Pillepich:2017}. One of the key ingredients that has allowed this progress is the inclusion of realistic models for the impact of feedback from the growth of super massive black holes \citep[e.g.][]{Bower:2017}. All successful models now demonstrate the need for active galactic nuclei (AGN) as an additional source of feedback that suppresses the formation of stars in high-mass haloes \citep[e.g.][]{Benson:2003,Croton:2006,Bower:2006,Crain:2015,Pillepich:2017}. One of the aims of the present paper is to compare the impact of the cosmological constant with that resulting from the inclusion of black holes (BHs) in the simulation. In a previous study, \cite{vandeVoort:2011} found that by preventing gas from accreting onto the central galaxies in massive haloes, outflows driven by AGN play a crucial role in the decline of the cosmic SFR. 

Different groups have used hydrodynamical simulations to study the effect of different dark energy or modified gravity models on cosmological, galactic and sub-galactic scales \citep[e.g.][]{Puchwein:2013,Penzo:2014,Penzo:2016}. Taking a different approach, in this paper we investigate the effect of the accelerated expansion of the Universe on galaxy formation by asking the following question:

\begin{quote}
\emph{How different would the Universe be if there had been no dark energy?}
\end{quote}

For our study, we use a suite of large hydrodynamical simulations from the \textit{Evolution and Assembly of GaLaxies and their Environment} (\textsc{eagle}) project \citep{Schaye:2015,Crain:2015}. Using state-of-the-art subgrid models for radiative cooling, star formation, stellar mass loss, and feedback from stars and accreting BHs, the simulations have reproduced many properties of the observed galaxy population and the intergalactic medium both at the present day and at earlier epochs \citep[e.g.][]{Furlong:2015-Sizes,Furlong:2015,Trayford:2015,Schaller:2015,Lagos:2015,Rahmati:2015,Rahmati:2015b, Bahe:2016, Rosas-Guevara:2016, Segers:2015}. Given that the physics of the real Universe is reasonably well captured by the phenomenological sub-grid models implemented in the simulations, with the use of appropriate assumptions, we can run the simulations beyond the present time, and explore the consequences of our models for the future. Furthermore, the simulation re-scaling strategy developed here, will be use in a companion paper \citep{Barnes:2018} considering a wider range of $\Lambda$ values, and determining the likelihood distribution of possible $\Lambda$ values conditioning the existence of observers.

The layout of this paper is as follows: In \Cref{sec:cosmo}, we develop a simple analytic model of the cosmic star formation rate that captures the suppression due to a cosmological constant. In \Cref{sec:Sim}, we briefly describe the simulations from which we derive our results and discuss our criteria for halo and galaxy definitions. In \Cref{sec:Sim} we also describe our motivations to run our cosmological simulations into the future, and our assumptions in doing so. \Cref{sec:scaling} provides a detailed discussion of our re-scaling strategy for the alternative cosmological models. In \Cref{sec:Results}, we explore the dependence of the star formation history of the universe on the existence of a cosmological constant and the presence of BHs. We also explore their impact on other galaxy population properties, both up to the present time, and into the future. Finally, we summarise and discuss our results in \Cref{sec:con}.

\begin{figure}\centering
\includegraphics[width=0.48\textwidth]{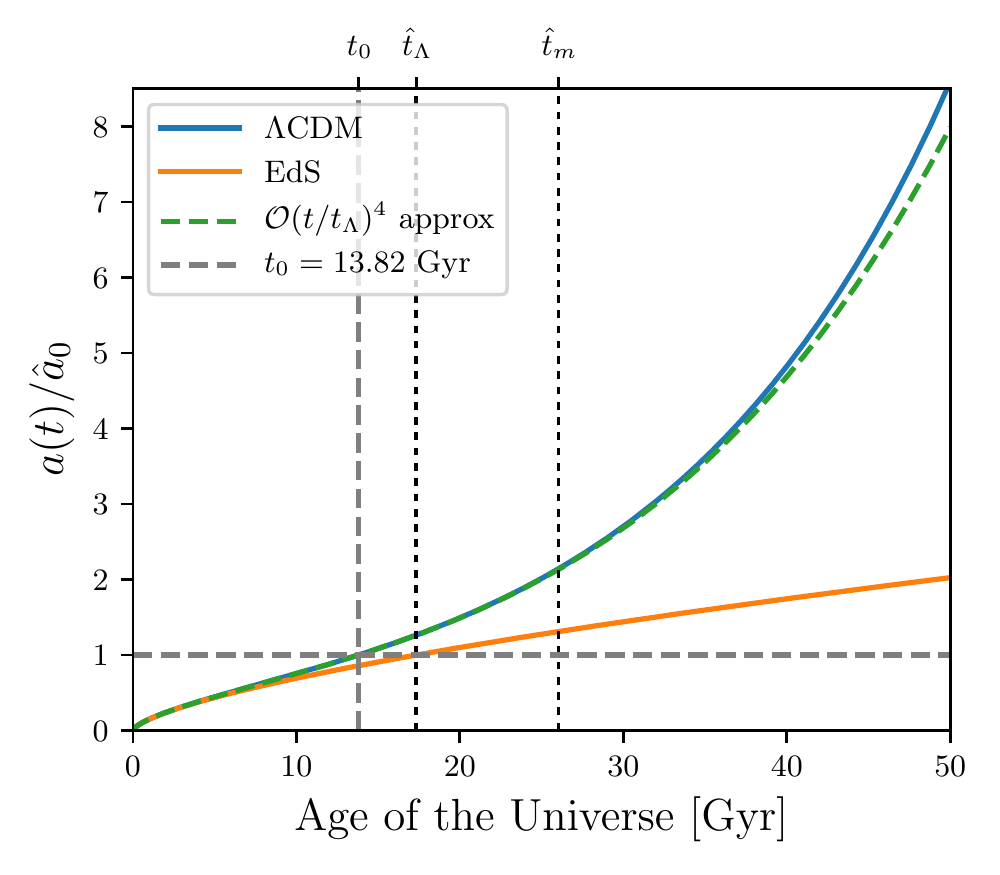}
 \vspace{-1.5em}
 \caption{Cosmic scale factor as a function of time for two cosmological models. The model for the cosmological parameters for a standard $\Lambda$CDM universe as inferred by the \citet{Planck} is shown in blue. An Einstein-de Sitter universe is shown in orange. Note that by construction the scale factors are indistinguishable when the universes are less than 1 Gyr old. The power series approximation of \cref{eq:a_of_t_power} is shown with a dashed green line.}
 \label{fig:a_t}
\end{figure}

\section{A simple analytic model for the cosmic star formation rate density}\label{sec:cosmo}

\subsection{Comparing different cosmological models}\label{sec:normalisation}
The star formation history of the Universe is determined by the interplay of cosmic expansion and the timescale at which cold gas can turn into stars. These processes happen on timescales that differ by several orders of magnitude, but are coupled through the accretion rate of gas onto gravitationally bound haloes. The aim of our paper is to compare theoretical universes in which the star formation timescales are the same, but the cosmological timescales vary. We need, therefore, to be careful when comparing the different models, since the choice of coordinates that vary with cosmological parameters will obscure the similarities of the models. In particular, the expansion factor at the present day, $a_0$, is often treated as an arbitrary positive number, and it is common practice to set $a_0 = 1$. In this paper, we need to take a different approach since we want to compare the properties of different universes at the same cosmic time (measured in seconds, or a multiple of key atomic transitions). Assuming a common inflationary origin, normalising out $a_0$ is not appropriate, since the expansion factor at the present day ($t_0=13.8$ Gyr), would be different for each universe. 
 
We still need to define a scale on which to measure the size of the universes we consider. Using a hat notation  $(\,\hat{}\,)$ to denote quantities in our observable Universe, we set $\hat{a}_0 \equiv \hat{a} (t_0) =1$. We want to emphasise that the cosmological models that we consider all start from very similar initial conditions. It therefore makes sense to normalise them to the same value of the expansion factor at an early time, $t_1$. We therefore set $a_1 \equiv a(t_1) = \hat{a}(t_1)$. We choose $\hat{a}_1 = 1/(1+127)$, corresponding to a redshift of $\hat{z}=127$ for a present-day observer in our Universe\footnote{$\hat{z}=127$ was the reference simulation's starting redshift.}. At this moment, the age of the universe is $t_1=11.98$~Myr. This applies to all of the universes we consider since the cosmological constant term has negligible impact on the expansion rate at such early times.

Although time (in seconds) is the fundamental coordinate that we use to compare universes, it is sometimes useful, for example when comparing to observational data, to express time in terms of the redshifts measured by a present-day observer in our Universe, $\hat{z}$. We convert between cosmic time $t$ (which is equivalent between universes) and $\hat{a}$ by inverting the 
time-redshift relation for our Universe:
\begin{equation}\label{eq:redshift}
	\hat{z} = \frac{\hat{a}_0}{\hat{a}(t)} - 1
\end{equation}
It is important to note that $\hat{z}$ is not the redshift that would be measured by an observer in an alternative universe.

In this paper, we will focus our comparison on two cosmological models, a standard $\Lambda$CDM universe as inferred by the \citet{Planck}, and an an Einstein deSitter (EdS) universe. Assuming both cosmological models have a common inflationary origin, the models can be normalised as follows:
\begin{enumerate}
	\item For the $\Lambda$CDM model (see \cref{tab:cosmo_params}) we set $\hat{a}_0 = \hat{a}(t_0)=1$, where $t_0=13.82$ Gyr is the present-day age of the universe. At time $t_1=11.98$~Myr, $\hat{a}_1 = \hat{a} (t_1) = 0.007813$. At this time, the expansion rate, as measured by the Hubble parameter is, $\hat{H}_1 = \hat{H} (t_1)= 54,377$ km/s/Mpc $= 55.6 \Gyr^{-1}$. 
	\item We require the EdS model to have the same early expansion history, i.e., $a(t_1) = 0.007813$ and $H(t_1) = 55.6 \Gyr^{-1}$. In this universe, at the present day (i.e. $t=t_0=13.82$ Gyr) the universe has a size, $a_0 = a(t_0) = 0.8589$ and an expansion rate, $H(t_0) = 0.0482 \Gyr^{-1} = 47.16$~km/s/Mpc. 
\end{enumerate}
 
\Cref{fig:a_t} shows the cosmic scale factor as a function of time for the two cosmological models. As expected, at $t=t_0$, an EdS universe is smaller in size at the present day, as the cosmic expansion has not been accelerated by the effect of $\Lambda$. As the two universes evolve into the future, the size differences and relative expansion rates grow, e.g. at $t=20\Gyr$, the scale factor for the $\Lambda$CDM models is ${\approx}25\%$ larger than for the EdS, and the expansion rate is ${\approx}50\%$ larger for our Universe.

\subsection{Cosmological expansion history as a function of time}

In the standard model of cosmology for a homogeneous and isotropic universe, the geometry of space-time is determined by the matter-energy content of the universe through the Einstein field equations as described by the Friedmann-Lema\^{i}tre-Robertson-Walker metric in terms of the scale factor $a(t)$ and the curvature $K$, yielding the well-known Friedmann equation,
\begin{equation}\label{eq:FRW}
\left(\frac{\dot{a}}{a}\right)^2 = H^2(t)= \frac{8 \pi G}{3} \rho - \frac{K c^2}{a^2} + \frac{\Lambda c^2}{3},
\end{equation}
where $H(t)$ is the Hubble parameter. As the inflationary models predict that the Universe should be spatially flat, we only consider universes with no spatial curvature, i.e. $K=0$. 

The density of \cref{eq:FRW} includes the contribution of non-relativistic matter and radiation ($\rho_{\mathrm{m}}$ and $\rho_{\mathrm{r}}$). The radiation content of the Universe dominated its global dynamics at very early times ($a \to 0$), but its contribution is negligible thereafter. Ignoring $\rho_{\mathrm{r}}$ and using the energy density at an arbitrary time $t_1$, \cref{eq:FRW} can be written as, 
\begin{equation}\label{eq:FRW2}
\left(\frac{\dot{a}}{a}\right)^2 = \frac{8 \pi G}{3} \rho_{\mathrm{m},1} \left(\frac{a}{a_1}\right)^{-3}   + \frac{\Lambda c^2}{3},
\end{equation}
where $\rho_{\mathrm{m},1}$ is the matter density of the universe at $t=t_1$, and $a_1 = a(t_1)$. We choose $t_1$ such that it corresponds to a sufficiently early epoch, when the contribution of the cosmological constant term is negligible. As discussed in the previous section, at this time any universe closely approximates an EdS universe and we can assume that  $a_1 = \hat{a}_1$ and $\rho_{\mathrm{m},1} =\hat{\rho}_{\mathrm{m},1} \implies  \hat{\rho}_{\mathrm{m},0} (\hat{a}_0 / \hat{a}_1)^3 = \rho_{\mathrm{m},0} (a_0 / a_1)^3$. Then, \cref{eq:FRW2} can be written as,
\begin{equation}\label{eq:FRW3}
\left(\frac{\dot{a}}{a}\right)^2 = \frac{8 \pi G}{3} \hat{\rho}_{\mathrm{m},0} \left(\frac{a}{\hat{a}_0}\right)^{-3}   + \frac{\Lambda c^2}{3}.
\end{equation}

Note that in \cref{eq:FRW3}, the evolution of the scale factor for any arbitrary cosmology is written in terms of the matter density of our Universe at the present time $\hat{\rho}_{\mathrm{m},0}$. We have left the factor of $\hat{a}_0$ explicit in the equation, but it can be set to $\hat{a}_0 = 1$, noting that ${a}_0 \neq 1$ for any cosmological model different to our Universe. 

The LHS of \cref{eq:FRW3}, has units of time$^{-2}$ and we will later find it useful to represent the RHS as the sum of two timescales.
The cosmological constant is often written as an energy component with energy density $\rho_\Lambda  = \Lambda c^2/8\pi G$,
however, we can express this as a timescale as follows,
 \begin{equation}
 	t_\Lambda = \sqrt{\frac{3}{\Lambda c^2}} = \frac{1}{H_0 \sqrt{\Omega_{\Lambda,0}}}.
 \end{equation}
Similarly, the matter content of the Universe can be expressed as a timescale,
 \begin{equation}
 	t_\mathrm{m} = \sqrt{\frac{3}{8\pi G\hat{\rho}_0}} = \frac{1}{\hat{H}_0 \sqrt{\hat{\Omega}_{\mathrm{m},0}}}.
 \end{equation}
Using the cosmological parameters for our Universe, $t_\Lambda=\hat{t}_\Lambda=17.33$ Gyr and $t_m = \hat{t}_m= 26.04$ Gyr. For an EdS universe, $t_\Lambda \rightarrow \infty$.

Using this notation, \cref{eq:FRW} can be written as,
\begin{equation}\label{eq:FRW_time}
\left(\frac{\dot{a}}{a}\right)^2 = t_m^{-2} \left(\frac{a}{\hat{a}_0}\right)^{-3} + t_\Lambda^{-2},
\end{equation}
which can be solved analytically to express the expansion factor as a function of time and the parameters $t_m$ and $t_\Lambda$:
\begin{equation}\label{eq:a_of_t}
a(t) = \left[\frac{1}{2} e^{-3{t}/{2t_\Lambda}} \left( e^{3{t}/{t_\Lambda}} -1 \right) \left(\frac{t_\Lambda}{t_m}\right)\right]^{2/3}
\end{equation}
In the limit $t_\Lambda \rightarrow \infty$ this reduces to the familiar EdS solution,
\begin{equation}
	\lim_{t_\Lambda\to\infty} a(t) = \left[\frac{3}{2} \frac{t}{t_m}\right]^{2/3}
\end{equation}

In order to explore the significance of the $t/t_\Lambda$ term 
 more clearly, we can expand \cref{eq:a_of_t} as a Taylor series:
\begin{equation}\label{eq:a_of_t_power}
a(t) \approx \left[\frac{3}{2} \frac{t}{t_m}\right]^{2/3}\left(1 + \frac{1}{4} \left(\frac{t}{t_\Lambda}\right)^2 + \frac{1}{80} \left(\frac{t}{t_\Lambda}\right)^4 +...\right)
\end{equation}
The coefficients of the series decreased rapidly so that the first three terms provide a good approximation up to $t = 2 t_\Lambda$ and beyond. \Cref{fig:a_t} shows how well this power series approximation works.

\begin{figure}\centering
\includegraphics[width=0.48\textwidth]{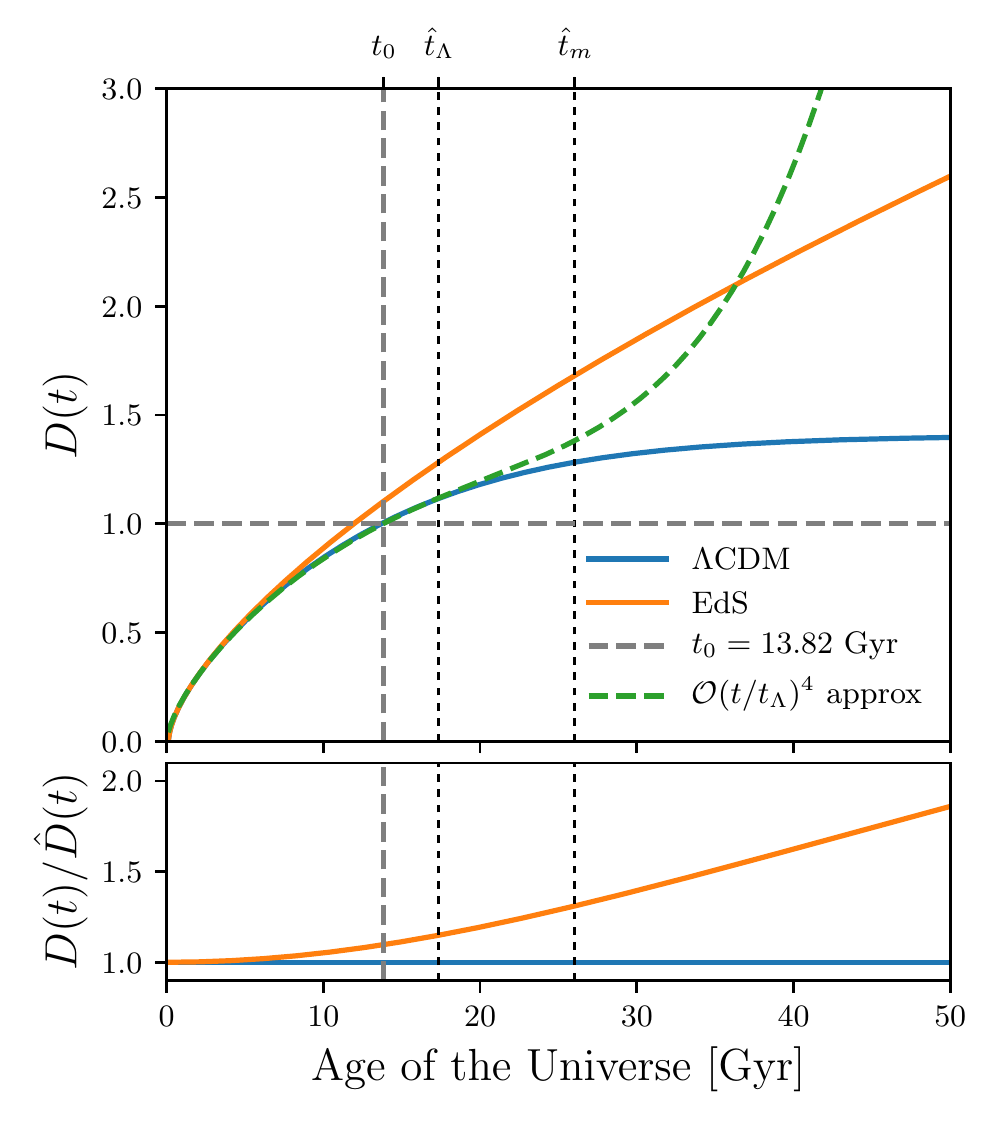}
 \vspace{-1.5em}
 \caption{The linear growth factor for the $\Lambda$CDM and EdS cosmological models. The rates are all normalised such that $\hat{D}(t_0) = 1$, for the $\Lambda$CDM model, and $D(t_1) = \hat{D}(t_1)$. The bottom panel shows the growth factor at a given time, divided by the growth factor for $\Lambda$CDM. The presence of a cosmological constant suppresses the growth of structure in the $\Lambda$CDM model (blue) compared to that in the EdS model (orange). The power series approximation of \cref{eq:D_of_t_approx} is shown with a dashed green line.} \label{fig:D_t}
\end{figure}

\begin{figure}
\centering
\includegraphics[width=0.48\textwidth]{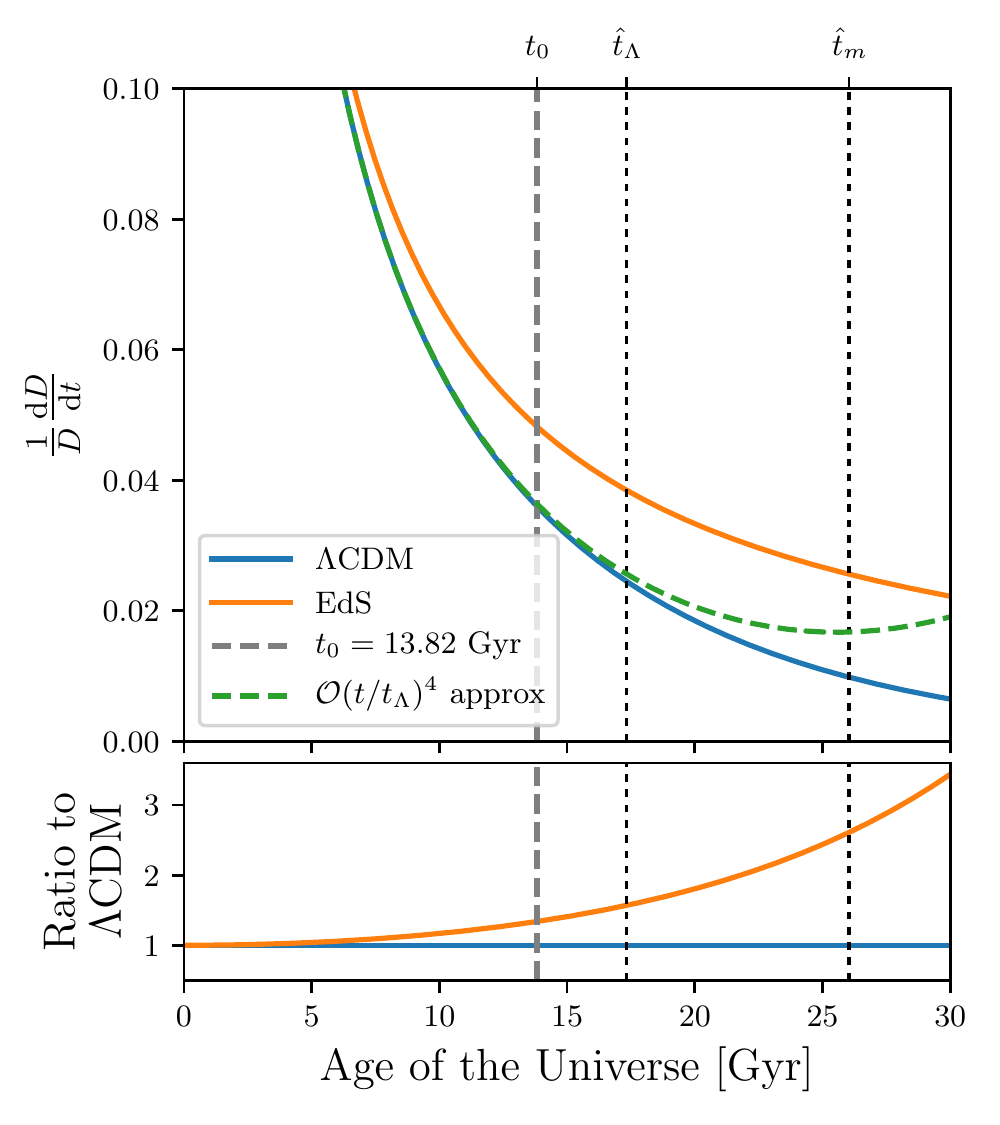}
 \vspace{-1.5em}
 \caption{The relative rate of growth of density perturbations, $\frac{1}{D}\frac{\dd D}{\dd t}$  for the $\Lambda$CDM and EdS cosmological models. The bottom panel shows the ratio, at a given time. The presence of a cosmological constant slows down the growth of structure in the $\Lambda$CDM model (blue) compared to that in the EdS model (orange). The power series approximation of \cref{eq:spec_dDdt} is shown with a dashed green line.}
 \label{fig:f_t}
\end{figure}

\subsection{The growth of density perturbations}

In the standard model of cosmology, structures such as galaxies and clusters of galaxies are assumed to have grown from small initial density perturbations. Expressing the density, $\rho$, in terms of the density perturbation contrast against a density background,
\begin{equation}
	\rho(\mathbf{x},t) = \bar{\rho}(t)[1+\delta(\mathbf{x},t)],
\end{equation}
the differential equation that governs the time dependence of the growth of linear perturbations in a pressureless fluid, such as e.g. dark matter, can be written as \citep[for a review see][]{Peebles:1980, Mo:2010},
\begin{equation}\label{eq:perturbations}
	\frac{\dd^2 \delta}{\dd t^2} + 2 \frac{\dot{a}}{a} \frac{\dd \delta}{\dd t} - 4\pi G \bar{\rho} \delta = 0.
\end{equation}

The growing mode of \cref{eq:perturbations} can be written as,
\begin{equation}
	\delta(t) = {D(t)} \delta(t_0),
\end{equation}
where $D(t)$ is the linear growth factor, which determines the normalisation of the linear matter power spectrum relative to the initial density perturbation power spectrum, and is computed by the integral 
\begin{equation}\label{eq:D_t}
D(t) \propto \frac{\dot{a}}{a} \int_0^t \frac{\dd t^\prime}{\dot{a}^2(t^\prime)}.
\end{equation}
Using the hat notation as before, we normalise $D(t)$ so that,
\begin{itemize}
	\item $\hat{D}_+(t_0) = 1$
	\item $D(t_1) = \hat{D}(t_1)$
\end{itemize}

In general, the growing mode can be obtained from \cref{eq:D_t} numerically. \Cref{fig:D_t} shows the growth factor as a function of cosmic time for the two cosmological models. As expected, the figure shows that linear perturbations grow faster in an EdS universe compared to those in a $\Lambda$CDM universe.

It is possible to gain more insight by integrating the power-series approximation for $a(t)$ from \cref{eq:a_of_t_power}. Expanding the solution again as a power series in $(t/t_\Lambda)$, retaining the leading terms, yields,  
\begin{equation}\label{eq:D_of_t_approx}
D(t) = \left[\frac{3}{2} \frac{t}{t_m}\right]^{2/3} \frac{2}{5} t_m^{2} K_D
            \left(1 - 0.1591 \left(\frac{t}{t_\Lambda}\right)^2 + 0.0366 \left(\frac{t}{t_\Lambda}\right)^4\right), 
\end{equation}
where $K_D$ is a normalisation constant. Requiring $\hat{D}(t_0) = 1$ gives $K_D = 4.70\times10^{-3} \Gyr^{-2}$. \Cref{fig:D_t} shows that \cref{eq:D_of_t_approx} provides a good approximation up to $t = t_\Lambda$.

This demonstrates that although the $t_\Lambda$ term slows down the growth of perturbations, its effect is less than 10\% until $t \sim  t_\Lambda\left( 0.1 / 0.1591 \right)^{1/2} \approx 0.8 t_\Lambda $ corresponding to $\approx 13.8$ Gyr (${\approx}\hat{t}_0$) in our Universe.  

As we discuss in the following section, the quantity of fundamental interest for the accretion rate of dark matter haloes is the relative rate of growth of density perturbations, $\frac{1}{D}\frac{dD}{dt}$.
We show this for the numerical solution in \cref{fig:f_t}. We can also compute the relative growth rate by differentiating the power-series approximation of \cref{eq:D_of_t_approx}. Retaining the lowest order terms, we find,
\begin{equation}\label{eq:spec_dDdt}
\frac{1}{D}\frac{\dd D}{\dd t} = \frac{2}{3t}\left(1 - 0.4773 \left(\frac{t}{t_\Lambda}\right)^2 + 0.1435\left(\frac{t}{t_\Lambda}\right)^4\right) 
\end{equation}
This expression does not depend on the constants $t_m$ or $K_D$ because we are focusing on the relative change in the growth factor. The impact of the cosmological constant term is relatively large, creating an ${\approx}50\%$ increase in growth rate for the EdS model compared to $\Lambda$CDM when $t\approx t_\Lambda$.

\begin{figure}\centering
\includegraphics[width=0.48\textwidth]{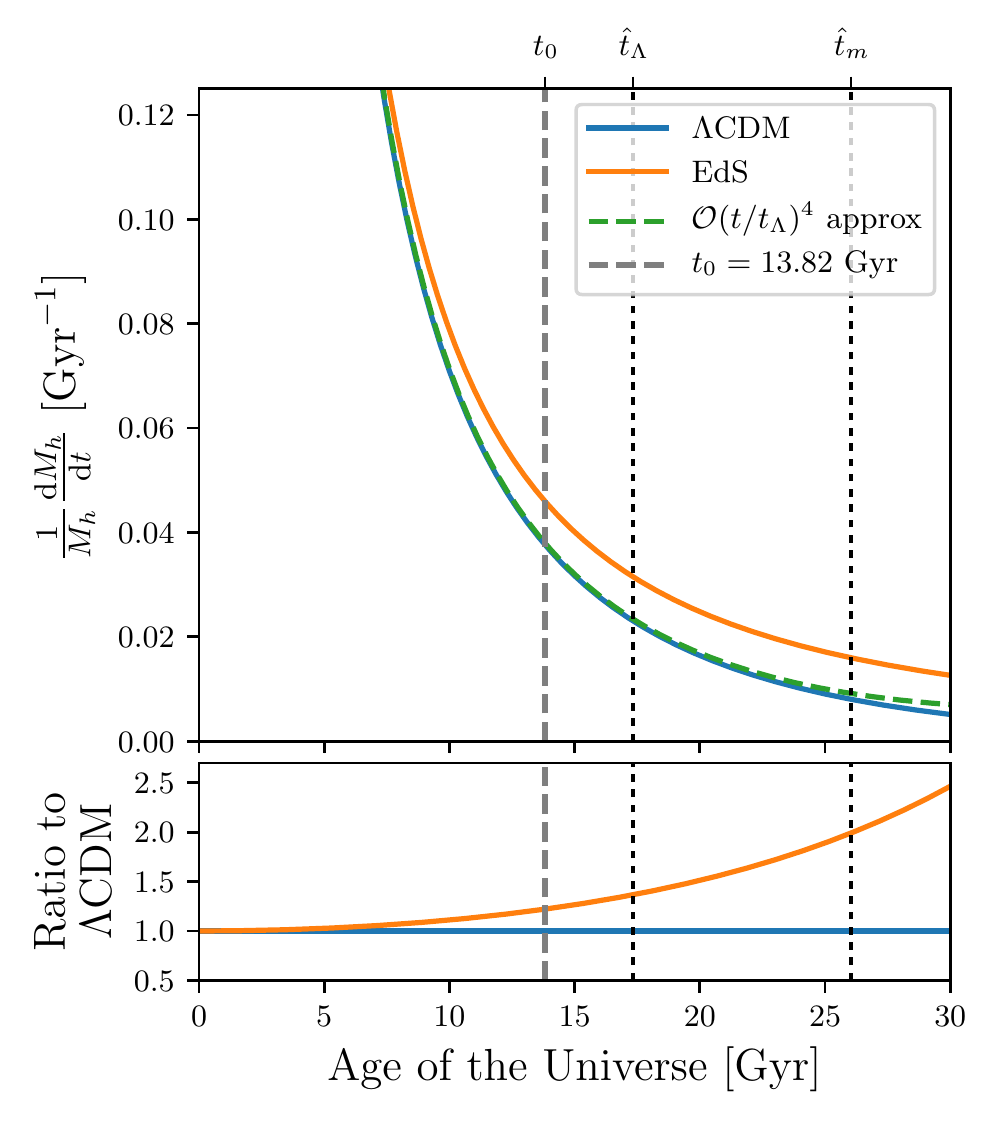}
 \vspace{-1.5em}
 \caption{The specific accretion rate of haloes of mass $M_h = 10^{12} \mathrm{M}_\odot$, $\frac{1}{M_h} \frac{\dd M_h}{\dd t}$ for the $\Lambda$CDM and EdS cosmological models. The bottom panel shows the ratio at a given time. The presence of a cosmological constant slows down the specific accretion rates of halos in the $\Lambda$CDM model (blue) compared to that in the EdS model (orange). The power series approximation of \cref{eq:mdot_eps_series} is shown with a dashed green line.}
 \label{fig:spec_growth}
\end{figure}

\subsection{Impact on halo accretion rates}

The growth rates of linear perturbations do not directly predict the growth rates of haloes, however, we can directly connect the two through the approach developed by Press \& Schechter \citep{Press_S:1974,Bond:1991,Bower:1991,Lacey_C:1993}. \cite{Correa:2015} showed that the accretion rates of haloes can be written as \citep[see also][]{Neistein:2006}, 
\begin{equation}\label{eq:mdot_eps}
\frac{1}{M_h} \frac{\dd M_h}{\dd t} = \sqrt{\frac{2}{\pi}}
     \frac{(\delta_c/D)}{S(M_h)^{1/2}\left(q^\gamma-1\right)^{1/2}}  
     \frac{1}{D}\frac{\dd D}{\dd t},
\end{equation}
where $M_h$ is the halo mass and $S(M_h)$ is the variance of the density field on the length scale corresponding the halo mass. $\delta_c$ is a parameter that represents a threshold in the linearly extrapolated density field for halo collapse. The parameters, $q$ and $\gamma$, are related to the shape of the power-spectrum around the halo mass $M_h$. Approximating the scale dependence of the density field as a power-law, $S=S_0 M_h^{-\gamma}$, \citealt{Correa:2015} find $S\approx 3.98, \gamma\approx 0.3$ and $q\approx 3.16$, giving $\left[S(M_h)\left(q^\gamma-1\right)\right]^{-1/2}\approx0.78$ for $10^{12} \Msol$ haloes. These values depend only on the initial power spectrum (which we assume to be the same in all the universes we consider) and do not depend on the cosmological parameters. This formulation thus neatly separates the contribution of the power-spectrum shape from the cosmological parameters. We are therefore able to assume that $q$ and $\gamma$ are the same for all the universes that we consider, and focus on the dependence on $D(t)$.

For the numerical values of the power-spectrum parameters around a halo mass of  $10^{12} \Msol$, Eq.~\ref{eq:mdot_eps} reduces to
\begin{equation}\label{eq:mdot_eps_numeric}
\frac{1}{M_h} \frac{\dd M_h}{\dd t} =  1.0456  \frac{1}{D^2} \frac{\dd D}{\dd t}. 
\end{equation}

This dependence can be understood as the combination of two factors. The first reflects the relative growth rate of density fluctuations $\frac{1}{D}\frac{\dd D}{\dd t}$. The second factor of $1/D$ comes from the rarity of haloes, reflecting the higher growth rate of fluctuations in the tail of the density field distribution.

Further insight can be gained by using the series approximation. This gives,
\begin{equation}\label{eq:mdot_eps_series}
\frac{1}{M_h} \frac{\dd M_h}{\dd t} = \frac{566.61}{\sqrt{S} \, t^{5/3}t_m^{4/3}}  \left( 1 - 0.3182  \left(\frac{t}{t_\Lambda}\right)^2 + 0.0563  \left(\frac{t}{t_\Lambda}\right)^4 \right).
\end{equation}
This explicitly shows how the presence of a cosmological constant modulates the halo growth rate. In our Universe, the impact of the cosmological constant term is relatively modest, however; at $t=13.8 \Gyr$, we expect the difference to be 20\%, growing to 40\% at $t=20\Gyr$.

As an example, in \cref{fig:spec_growth} we show the accretion rate of haloes of $M_h = 10^{12} \mathrm{M}_\odot$, both numerically and using \cref{eq:mdot_eps_series}.

\subsection{Impact on the star formation rate of the Universe}\label{sec:approx}

In order to link the SFR of halos of mass $M_h$ to their accretion rate, as a first approximation, we assume a time-independent galaxy specific star formation rate to host halo specific mass accretion rate relation \citep[e.g.][]{Behroozi:2013_Evol,Tacchella:2013,Rodriguez:2016},
\begin{equation}
	\frac{\dot{M}_*/{M}_*}{\dot{M}_h/{M}_h} = \frac{\partial \mathrm{log} M_*}{\partial \mathrm{log} M_h} = \epsilon(M_h),
\end{equation}
where the star formation efficiency $\epsilon$, of haloes of mass $M_h$, is the slope of the stellar-halo mass relation. From this equation, the star formation as a function of halo mass can be written as,
\begin{equation}
	\dot{M}_*(M_h) = \epsilon_*(M_h) \dot{M}_h,
\end{equation}
where $\epsilon_*(M_h) := \epsilon(M_h) \times (M_*/M_h)$ is completely defined by the stellar-halo mass relation. As there is no a priori knowledge of the functional form of $\epsilon_*(M_h)$, we use the abundance matching results from \citealt{Behroozi:2013} to estimate $\epsilon_*(M_h)$. The efficiency $\epsilon_*(M_h)$ peaks at masses similar to Milky-Way sized halos (${\sim} 10^{12} \Msol$) and falls steeply for higher and lower masses. $\epsilon_*(M_h)$ can be  well approximated by a broken power law as, 
\begin{equation}\label{eq:epsilon}
\epsilon_*({M_h})\propto
\begin{cases}
	\left(\frac{M_h}{10^{12}\mathrm{M}_\odot}\right)^1 & \text{if $M_h \leq 10^{12}\mathrm{M}_\odot $} \\ \\
	\left(\frac{M_h}{10^{12}\mathrm{M}_\odot}\right)^{-1} & \text{if $M_h > 10^{12}\mathrm{M}_\odot $} \\
\end{cases}	
\end{equation}
At low masses, star formation rate is suppressed because of the efficiency of feedback from star formation, at higher masses the cooling of the inflowing gas is suppressed by heating from black holes \citep{WF:1991,Benson:2003,Bower:2006,Haas:2013,Crain:2015,Dubois:2016,Bower:2017}.

In order to complete the analysis, we need to combine the specific halo mass accretion rate with an estimate of the halo abundance. 
\begin{figure}\centering \includegraphics[width=0.48\textwidth]{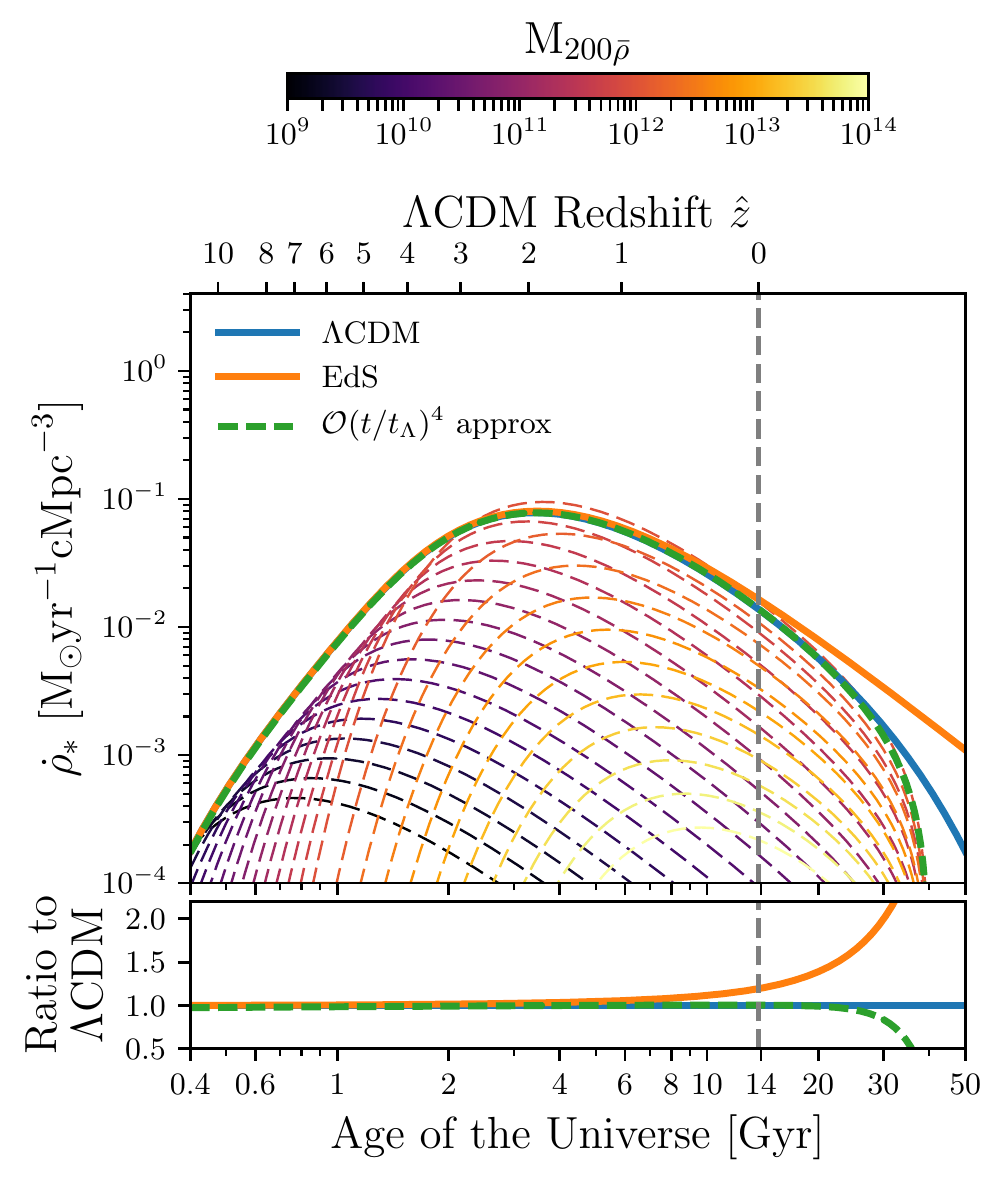}
 \vspace{-1.5em}
 \caption{The predicted SFR history of the Universe, and the expected influence of the cosmological constant using the simple model developed in \Cref{sec:approx}.  Coloured lines show the contributions from dark matter haloes of different masses (per dex), using the star formation efficiency described by \cref{eq:ssfr_rate_density_dm}. The total SFR for the $\Lambda$CDM universe calculated numerically is shown in blue. An Einstein-de Sitter universe is shown in orange. The integrated SFR calculated using the approximation of \cref{eq:ssfr_rate_density_dm,eq:rho_int}, is shown with a dashed green line. The bottom panel shows the ratio at a given time. The predicted suppression of SFR due to $\Lambda$ at the present time is ${\approx}19\%$. At $t\approx30$ Gys the predicted SFR density for the EdS model is double than $\Lambda$CDM, and ${\approx}6$ times higher at $t=50$ Gyr. The approximation of \cref{eq:ssfr_rate_density_dm,eq:rho_int} ceases to work for $t \gtrsim 25$ Gyr.}
 \label{fig:madau_prediction}
\end{figure}

In the Press \& Schechter analysis, the co-moving abundance of haloes of mass $M_h$ at time $t$ is given by \citep{Press_S:1974,Bond:1991,Bower:1991,Lacey_C:1993},
\begin{equation}\label{eq:dndm_eps}
\frac{dn(M_h,t)}{dM_h} = \frac{\hat{\rho}_0}{M_h^2} \frac{\delta_c \gamma}{\sqrt{2\pi} S^{1/2}} \frac{1}{D}
         \exp\left( - \frac{\delta_c^2}{2 S D^2}\right)  
\end{equation}
where we have assumed that the density power spectrum is a power law with exponent $\gamma$ and written the co-moving density of the Universe as $\hat{\rho}_0$ following our convention. Note that we compute co-moving densities. At the same cosmic time, the different expansion rates will result in different physical (proper) halo and SFR densities, simply because of the more rapid expansion of the $\Lambda$CDM cosmology.  

The total cosmic SFR density is given by the integral of all star formation in all haloes,
\begin{equation}\label{eq:rho_int}
	\dot{\rho}_*(t) = \int \dot{M}_*(M_h) \frac{\dd n(M_h,t)}{\dd M_h} \dd M_h = \int \epsilon_*(M_h) \dot{M}_h \frac{\dd n(M_h,t)}{\dd M_h} \dd M_h
\end{equation}

Using the power series approximation \cref{eq:mdot_eps_series} together with \cref{eq:epsilon} and \cref{eq:dndm_eps}, the contribution to the cosmic SFR density from haloes of mass $M_h$ (the integrand of \cref{eq:rho_int}) is given by,
\begin{equation}\label{eq:ssfr_rate_density_dm}
	\begin{aligned}
		\frac{\dd \dot{\rho_*}}{\dd M_h} & = \epsilon_*(M_h)\left[ \frac{1}{M_h} \frac{\dd M_h}{\dd t} \right] M_h \frac{\dd n(M_h,t)}{\dd M_h} \\ 
		 &= \epsilon_*(M_h) \frac{46230.9 \hat{\rho}_0}{M_h S t^{7/3} t_m^{8/3}} \left(1 - 0.1590 \left(\frac{t}{t_\Lambda}\right)^2 - 0.0056 \left(\frac{t}{t_\Lambda}\right)^4 \right) \\
		 & \times \exp \left[ -\frac{232382}{S \, t^{4/3} t_m^{8/3}} \left(1 + 0.3182 \left(\frac{t}{t_\Lambda}\right)^2 + 0.0028 \left(\frac{t}{t_\Lambda}\right)^4\right)\right].
	\end{aligned}
\end{equation}

The cosmological constant term enters through both the multiplier and the exponential terms, with a balance that depends on the halo mass through $S$ (see \cref{eq:mdot_eps}). While smaller haloes are more abundant than large objects, a smaller fraction of the inflowing material is converted into stars. As a result, the SFR  density is dominated by the largest haloes in which star formation is able to proceed without generating efficient BH feedback.
The smaller haloes only contribute significantly at very early times, when the abundance of larger objects is strongly suppressed by the exponential term. We see therefore that the level of suppression expected for ${\approx}10^{12}\Msol$ haloes is representative of most of the SFR in the Universe.

The predictions for the contributions of different halo masses are shown in \cref{fig:madau_prediction}, together with the total expected cosmic SFR density, for the two cosmologies that we consider in this paper. We will compare this approximation in \cref{sec:Results} to the results from the \textsc{eagle} simulations. 

\begin{figure*}
\includegraphics[width=0.95\textwidth]{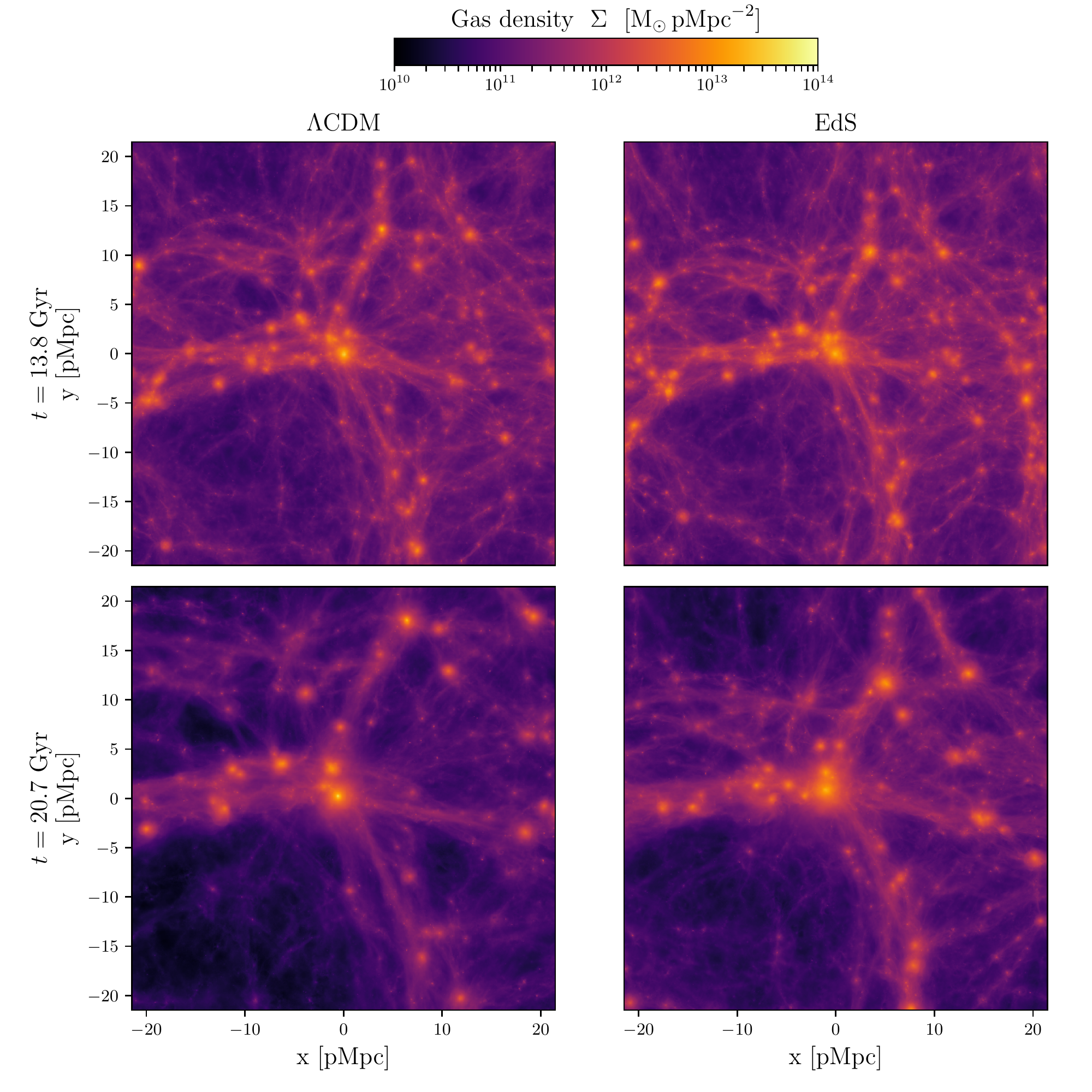}
 \caption{The evolution of the projected gas density for each \textsc{eagle} model  centred on the most massive halo at the present time ($t = 13.8$ Gyr). The length of each image is $43$ (proper) Mpc on a side, to highlight the difference on cosmic expansion. Left: $\Lambda$CDM universe. Right: EdS universe. Top: Cosmic time $t = 13.8$ Gyr. Bottom: Cosmic time $t = 20.7$ Gyr. The colour coding represents the (proper) surface gas density projected along the line of sight. At $t = 13.8$ Gyr, the general appearance of both models is similar, as the phases of the initial fluctuations are the same. Over the next $6.8$ Gyr, the effect of $\Lambda$ becomes more significant, slowing down the growth of structure compared to the EdS model.}
 \label{fig:img}
\end{figure*}

\section{The EAGLE Simulations}\label{sec:Sim}

The simple analytic model provides a basis for interpreting the results, but it is highly simplified. We therefore compare the analytic model to numerical hydrodynamic simulations based on the \textsc{eagle} project. The \textsc{eagle} simulation suite\footnote{\url{http://www.eaglesim.org}} \citep{Schaye:2015,Crain:2015} consists of a large number of cosmological hydrodynamical simulations that include different resolutions, simulated volumes and physical models. These simulations use advanced smoothed particle hydrodynamics (SPH) and state-of-the-art subgrid models to capture the unresolved physics. The simulation suite was run with a modified version of the \textsc{gadget-}${\scriptstyle 3}$  SPH code (last described by \citealt{GADGET}) and includes a full treatment of gravity and hydrodynamics. The calibration strategy is described in detail by \cite{Crain:2015} who also presented additional simulations to demonstrate the effect of parameter variations. 

The halo and galaxy catalogues for more than $10^5$ simulated galaxies of the main \textsc{eagle} simulations with integrated quantities describing the galaxies, such as stellar mass, SFRs, metallicities and luminosities, are available in the \textsc{eagle} database\footnote{\url{http://www.eaglesim.org/database.php}} \citep{McAlpine:2015-DB}. A complete description of the code and physical parameters used can be found in \citet{Schaye:2015}. 

The \textsc{eagle} reference simulations used cosmological parameters measured by the \citet{Planck}. In this paper we introduce three main \textsc{eagle} simulations that use the same calibrated sub-grid parameters as the reference model, but change the cosmological model by setting the cosmological constant to zero, and/or removing feedback from BHs. The values of the cosmological parameters used for the  simulations are listed in \cref{tab:cosmo_params}. The values of other relevant parameters adopted by all simulations featured in this study are listed in \cref{tab:a}. Together these parameters determine the dynamic range and resolution that can be achieved by the simulations.

\begin{table}
\centering
\caption{The cosmological parameters for the \textsc{eagle} simulations used in this study. $\Lambda$CDM model refers to parameters inferred by the \citet{Planck}. EdS refers to an Einstein-de Sitter universe. $\Omega_\mathrm{m}, \Omega_\Lambda, \Omega_\mathrm{b}$ are the average densities of matter, dark energy, and baryonic matter in units of the critical density at redshift zero; $H_0$ is the Hubble constant, $\sigma_8(t_1)$ is the square root of the linear variance of the matter distribution at the initial cosmic time of the simulations ($t_1=11.98 \,\mathrm{Myr}$) when smoothed with a top-hat filter of radius $11.8$ cMpc ($8 \, h^{-1}$ cMpc for a $\Lambda$CDM model), $n_s$ is the scalar power-law index of the power spectrum of primordial adiabatic perturbations, and $Y$ is the primordial abundance of helium. Values in bold show differences with respect to the $\Lambda$CDM values.} \label{tab:cosmo_params}
\begin{tabular}{lll}
\hline
\multicolumn{1}{c}{Cosmological Parameter}                                & $\Lambda$CDM (Ref) & EdS     \\ \hline
$\Omega_\mathrm{m}$                                                       & 0.307        & \textbf{1}       \\
$\Omega_\Lambda$                                                          & 0.693        & \textbf{0}       \\
$\Omega_\mathrm{b}$                                                       & 0.04825      & \textbf{0.15717} \\
$h \equiv H_0/(100 \, \mathrm{km}\,\mathrm{s}^{-1} \, \mathrm{Mpc}^{-1})$ & 0.6777       & \textbf{0.3754}  \\
$\sigma_8(t_1)$                                                                & 0.0083       & 0.0083  \\
$n_s$                                                                     & 0.9611       & 0.9611  \\
$Y$                                                                       & 0.248        & 0.248  \\
\hline
\end{tabular}
\end{table}

\begin{table*}
\caption{Box-size, number of particles, initial baryonic and dark matter particle mass, co-moving and Plummer-equivalent gravitational softening, inclusion of AGN feedback, cosmological model and Hubble parameter for the \textsc{eagle} simulations used in this paper. Values in bold show differences with respect to the Ref simulation. The three bottom small box models were used for convergence tests.}\label{tab:a}
\begin{tabular}{cccccccccc}\hline
\multicolumn{1}{l}{Identifier} & L & N & ${m}_{\mathrm{gas}}$ & ${m}_{\mathrm{DM}}$ & $\epsilon_{\mathrm{com}}$ &    $\epsilon_{\mathrm{prop}}$ & AGN & Cosmology & h\\
 & $[\mathrm{cMpc}]$ &  & $[\mathrm{M}_{\sun}]$  & $[\mathrm{M}_{\sun}]$ & $[\mathrm{ckpc}]$ & $[\mathrm{pkpc}]$ & &\\ \hline
\multicolumn{1}{l}{$\Lambda$CDM (Ref)} & ${50}$ & \multicolumn{1}{r}{$2 \times 752^3$} & $1.81 \times 10^6$ & $9.70 \times 10^6$ & $2.66$ & $0.70$ & Yes & Planck 14 & 0.6777\\
\multicolumn{1}{l}{$\Lambda$CDM (No AGN)} & ${50}$ & \multicolumn{1}{r}{$2 \times 752^3$} & $1.81 \times 10^6$ & $9.70 \times 10^6$ & $2.66$ & $0.70$ & \textbf{No} & Planck 14 & 0.6777\\
\multicolumn{1}{l}{EdS} & ${50}$ & \multicolumn{1}{r}{$2 \times 752^3$} & $1.81 \times 10^6$ & $9.70 \times 10^6$ & $2.66$ & $0.70$ & Yes & \textbf{EdS} & \textbf{0.3754}\\
\multicolumn{1}{l}{EdS (No AGN)} & ${50}$ & \multicolumn{1}{r}{$2 \times 752^3$} & $1.81 \times 10^6$ & $9.70 \times 10^6$ & $2.66$ & $0.70$ & \textbf{No} & \textbf{EdS} & \textbf{0.3754}\\ \hdashline

\multicolumn{1}{l}{$\Lambda = 0$ L12 h0\_3754} & ${12.50}$ & \multicolumn{1}{r}{$2 \times 188^3$} & $1.81 \times 10^6$ & $9.70 \times 10^6$ & $2.66$ & $0.70$ & Yes & \textbf{EdS} & \textbf{0.3754}\\
\multicolumn{1}{l}{$\Lambda = 0$ L12 h0\_6777} & ${8.43}$ & \multicolumn{1}{r}{$2 \times 188^3$} & $1.81 \times 10^6$ & $9.70 \times 10^6$ & $\mathbf{1.79}$ & $0.70$ & Yes & \textbf{EdS} & 0.6777\\
\multicolumn{1}{l}{$\Lambda = 0$ L12 h0\_4716} & ${10.73}$ & \multicolumn{1}{r}{$2 \times 188^3$} & $1.81 \times 10^6$ & $9.70 \times 10^6$ & $\mathbf{2.28}$ & $0.70$ & Yes & \textbf{EdS} & \textbf{0.4716}\\ \hline
\end{tabular}
\end{table*}

\cref{fig:img} Shows the projected gas density for the $\Lambda$CDM and EdS cosmological models both at the present day  and into the future. At $t = 13.8$ Gyr, the general appearance of both models is similar, but over the next $6.8$ Gyr,  the effect of $\Lambda$ becomes more significant slowing down the growth of structure.

\subsection{Subgrid models}\label{sec:subgrid}

Processes that are not resolved by the simulations are implemented as subgrid physical models; they depend solely on local interstellar medium (ISM) properties. A full description of these subgrid models can be found in \cite{Schaye:2015}. In summary:
\begin{enumerate}
	\item Radiative cooling and photoheating are implemented element-by-element as in \cite{Wiersma:2009Cooling}, including the 11 elements found to be important, namely, H, He, C, N, O, Ne, Mg, Si, S, Ca, and Fe. Hydrogen reionization is implemented by switching on the full \cite{HM} background at the proper time corresponding to redshift $z=11.5$ in our $\Lambda$CDM Universe.
	\item Star formation is implemented stochastically following the pressure-dependent Kennicutt-Schmidt relation as in \cite{SchayeDallaVecchia:2008}. Above a metallicity-dependent density threshold $n^∗_\mathrm{H}(Z)$, which is designed to track the transition from a warm atomic to an unresolved, cold molecular gas phase \citep{Schaye:2004}, gas particles have a probability of forming stars determined by their pressure.
	\item Time-dependent stellar mass loss due to winds from massive stars and AGB stars, core collapse supernovae and type Ia supernovae, is tracked following \cite{Wiersma:2009Enrichment}. 
	\item Stellar feedback is treated stochastically, using the thermal injection method described in \cite{DallaVecchiaSchaye:2012}. 
	\item Seed BHs of mass $\mathrm{M} = 1.48 \times 10^5 \mathrm{M}_\odot$, are placed in haloes with a mass greater than $1.48 \times 10^{10} \mathrm{M}_\odot$ and tracked following the methodology of \cite{SpringelDiMatteoHernquist:2005,Booth:2009}. Accretion onto BHs follows a modified version of the Bondi-Hoyle accretion rate which takes into account the circularisation and subsequent viscous transport of infalling material, limited by the Eddington rate as described by \cite{Rosas-Guevara:2015}\footnote{The \textsc{eagle} simulation do not include a boost factor the accretion rate of BHs to account for an unresolved clumping factor.}. Additionally, BHs can grow by merging with other BHs as described in \cite{Schaye:2015,Salcido}.
	\item Feedback from AGN is implemented following the stochastic heating scheme described by \cite{Schaye:2015}. Similar to the supernova feedback, a fraction of the accreted gas onto the BH is released as thermal energy with a fixed heating temperature into the surrounding gas following \cite{Booth:2009}.
	\end{enumerate}
	
For the \textsc{eagle} simulations, the subgrid parameters were calibrated to reproduce three properties of galaxies at redshift $z = 0$: the galaxy stellar mass function, the galaxy size--stellar mass relation, and the black hole mass-stellar mass relation\footnote{BH feedback efficiency left unchanged from \cite{Booth:2009}.}. The calibration strategy is described in detail by \cite{Crain:2015}, who explores the effect of parameter variations.

\subsection{Halo and galaxy definition}

Haloes were identified running the ``Friends-of-Friends'' (FoF) halo finder on the dark matter distribution, with a linking length equal to 0.2 times the mean inter-particle spacing. Galaxies were identified as self-bound over-densities within the FoF group using the \textsc{subfind} algorithm \citep{Springel:2001,Dolag:2009}. A `central' galaxy is the substructure with the largest mass within a halo. All other substructures within a halo are `satellite' galaxies.

Comparing haloes from simulations with different cosmologies is not a well-defined task, as halo masses are usually defined in terms of quantities that depend on the specific cosmological parameters. Typically, this is done by growing a sphere outwards from the potential minimum of the dominant dark matter sub-halo out to a radius where the mean interior density equals a fixed multiple of the critical or mean density of the Universe, causing an artificial `pseudo-evolution' of dark matter halos by changing the radius of the halo \citep{Diemer:2013}. Star formation, however, is governed by the amount of gas that enters these halos and reaches their central regions. \citealt{Wetzel:2015} show that the growth of dark matter haloes is subject to this `pseudo-evolution', whereas the accretion of gas is not. Because gas is able to cool radiatively, it decouples from dark matter, tracking the accretion rate near a radius of $R_{200{\bar{\rho}}}$, the radius within which the mean density is 200 times the mean density of the universe, $\bar{\rho}$. As we try to connect the accretion of dark matter haloes to star formation, we define halo masses as the total mass within $R_{200{\bar{\rho}}}$,
\begin{equation}
	M_{200_{\bar{\rho}}} = 200 \frac{4\pi}{3} R^3_{200{\bar{\rho}}} \bar{\rho}.
\end{equation}
Additionally, as $\bar{\rho} = \Omega_{m}(t) \rho_{c}(t)$ is given in co-moving coordinates, the mean density of the universe remains constant in time for each cosmological model. 

Following \citealt{Schaye:2015} and \citealt{Furlong:2015}, galaxy stellar masses are defined as the stellar mass associated with the subhalo within a 3D 30 proper kilo parsec (pkpc) radius, centred on the minimum of the subhalo's centre of gravitational potential. This definition is equivalent to the total subhalo mass for low-mass objects, but excludes diffuse mass around very large subhaloes, which would contribute to the intracluster light (ICL).

\subsection{Continuing the simulations into the future}\label{sec:future}

As $\Lambda$ continues driving the accelerated expansion of the universe, the linear growth of density perturbations, $D(t)$ is suppressed (see \cref{eq:D_of_t_approx}). Further insight can be obtained if we analyse the evolution of the potential perturbations given by the perturbed Poisson equation for an expanding space,
\begin{equation}
	\nabla^2 \Phi = 4 \pi G \bar{\rho}_{\mbox{\tiny p}} a^2 D\delta_0, 
\end{equation}
where the Laplace operator is with respect to comoving coordinates, and the mean density $\bar{\rho}_{\mbox{\tiny p}}$ is given in proper coordinates. As $\bar{\rho}_{\mbox{\tiny p}}$ evolves $\propto a^{-3}$, it follows that $\nabla^2 \Phi \propto D/a$. Using \cref{eq:a_of_t_power} and \cref{eq:D_of_t_approx} we can see that for an EdS universe, both $D$ and $a$ are $\propto t^{2/3}$ and the potentials are expected to stop evolving (they are frozen in). On the other hand, the suppression of growth of density perturbations due to a cosmological constant causes a decay in the potentials as the universe expands. As shown in \cref{fig:D_t}, according to linear theory, these two scenarios have comparable growth factors at the present time (${\approx}10 \%$ difference, see Eq.~\ref{eq:D_of_t_approx}), but the difference becomes increasingly important in the future. Furthermore, star formation is expected to eventually exhaust the finite reservoir of cold gas in galaxies, shutting off the production of stars in the universe forever \citep[e.g.][]{Fukugita:1998,Loeb:2016}. 

In order to study the impact of $\Lambda$ in galaxy formation beyond the present day, and hence explore the uniqueness of the present epoch, and in order to determine the total mass of stars ever produced by the universe, we allow the simulations to run into the future, i.e. $t>t_0$ \cite[e.g.][]{Barnes:2005,Loeb:2016}. The subgrid models for star formation, stellar mass loss, stellar feedback, BH seeding and feedback from AGN were kept as described in \cref{sec:subgrid} as the simulations ran into the future. On the other hand, as there is no information about the UV and X-ray background radiation from quasars and galaxies into the future, for simplicity, we assumed that the background radiation freezes out, i.e. we kept its value at $t=t_0$ constant into the future. We consider this to be a good simplification as the UV background only affects star formation in very low mass haloes, and hence does not affect the cosmic SFR at late times \citep[e.g.][]{Schaye_et_al:2010}. 

\begin{table}
\centering
\caption{Parameters re-scaled in the initial conditions. Hat notation indicates parameters for our Universe.}\label{tab:ICs}
\begin{tabular}{lcc} \hline
\multicolumn{1}{c}{Parameter} & Units                     & Re-scaling factor                                                                                              \\ \hline
Box size                        & $\mathrm{cMpc} \, h^{-1}$    & $(\hat{h}^{-1} h) \times (\hat{a}_{1} a_{1}^{-1})$ \\
Particle Masses                 & $\mathrm{M}_\odot \, h^{-1}$ & $(\hat{h}^{-1} h)$                                                                        \\
Particle Coordinates            & $\mathrm{cMpc} \, h^{-1}$    & $(\hat{h}^{-1} h) \times (\hat{a}_{1} a_{1}^{-1})$ \\
Particle Velocities             & $\mathrm{cMpc} \, s^{-1}$    & $ (\hat{a}_{1} a_{1}^{-1}) ^{1/2}$  \\ \hline                                   
\end{tabular}
\end{table}

\begin{table}
\centering
\caption{Additional parameters re-scaled in the simulations. Hat notation indicates parameters for our Universe.}\label{tab:parameters}
\begin{tabular}{lcc} \hline
\multicolumn{1}{c}{Parameter} & Units                     & Re-scaling factor                                                                                              \\ \hline
Co-moving Softening         			    & $\mathrm{ckpc} \, h^{-1}$    & $(\hat{h}^{-1} h) \times (\hat{a}_{1} a_{1}^{-1})$ \\
Max Softening                			    & $\mathrm{pkpc} \, h^{-1}$     & $(\hat{h}^{-1} h)$                                                                        \\
Seed BH Mass 			                    & $\mathrm{M}_\odot \, h^{-1}$ & $(\hat{h}^{-1} h)$                                                                        \\
Min $\mathrm{M}_\mathrm{FOF}$ for New BH	& $\mathrm{M}_\odot \, h^{-1}$ & $(\hat{h}^{-1} h)$                                                                        \\ \hline
\end{tabular}
\end{table}

\section{Simulations re-scaling}\label{sec:scaling}

In this section we describe our simulation re-scaling strategy. At early epochs, the universe was matter dominated, and so we can neglect the contribution of $\Lambda$. Hence, any universe with non zero matter density, i.e. $\rho_{\mathrm{m},0} \neq 0$, will be close to an EdS universe at early epochs. Therefore, we can assume identical initial conditions for all cosmological models of interest here.

The initial conditions for the reference $\Lambda$CDM model were created in three steps. First, a particle load, representing an unperturbed homogeneous periodic universe was produced. Secondly, a realisation of a Gaussian random density field with the appropriate linear power spectrum was created over the periodic volume. Thirdly, the displacements and velocities, consistent with the pure growing mode of gravitational instability, were calculated from the Gaussian realisation and applied to the particle load producing the initial conditions. The initial density perturbation power spectrum is commonly assumed to be a power-law, i.e. $P_i(k) \propto k^{n_s}$. From the Planck results \citep{Planck}, the spectral index $n_s$, has a value of $n_s = 0.9611$. A transfer function with the cosmological parameters shown in \cref{tab:cosmo_params} was generated using CAMB (version Jan\_12; \citealt{CAMB}). The linear matter power spectrum was generated by multiplying the initial power spectrum by the square of the dark matter transfer function evaluated at the present day $t=t_0$, i.e. $P(k,t) = P_i(k)T^2(k)D^2(t)$.\footnote{The CAMB input parameter file and the linear power spectrum are available at \url{http://eagle.strw.leidenuniv.nl/}}

The \textsc{eagle} version of \textsc{gadget} uses an internal system of units that includes both co-moving coordinates and the dimensionless Hubble parameter, $h$. For the alternative cosmological models, we have the freedom to choose the present time $t_0$ for each simulation, and we re-scale all the initial condition such that they are identical in physical \textit{``h-free''} units at an early time $t_1=11.98$~Myr. \Cref{tab:ICs} shows the parameters that have been re-scaled in the initial conditions. 

The same tables of radiative cooling and photoheating rates as a function of density and temperature were used for all cosmological models. The corresponding redshifts for the cooling tables were re-scaled such that they correspond to the same cosmic time for each cosmology. That is, using \cref{eq:a_of_t}, we find the scale factor, $a$, for which the alternative cosmology satisfies, 
\begin{equation}
	{t} (\hat{a}) = t (a).
\end{equation}

 The average baryonic density $\Omega_\mathrm{b}$ has been re-scaled in such way that the baryon fraction ($f_\mathrm{b} = \Omega_\mathrm{b}/\Omega_\mathrm{m}$) is equal in both cosmologies, i.e.
 \begin{equation}
 	\frac{\hat{\Omega}_\mathrm{b}}{\hat{\Omega}_\mathrm{m}} = \frac{\Omega_\mathrm{b}}{\Omega_\mathrm{m}}.
 \end{equation}
 
 \Cref{tab:parameters} shows additional parameters that have been re-scaled to be equivalent in $h$-free physical units. Finally, hydrogen and He\textsc{ii} reionization were also re-scaled in such way that redshifts correspond to the same cosmic time.

\begin{figure}
 \includegraphics[width=0.48\textwidth]{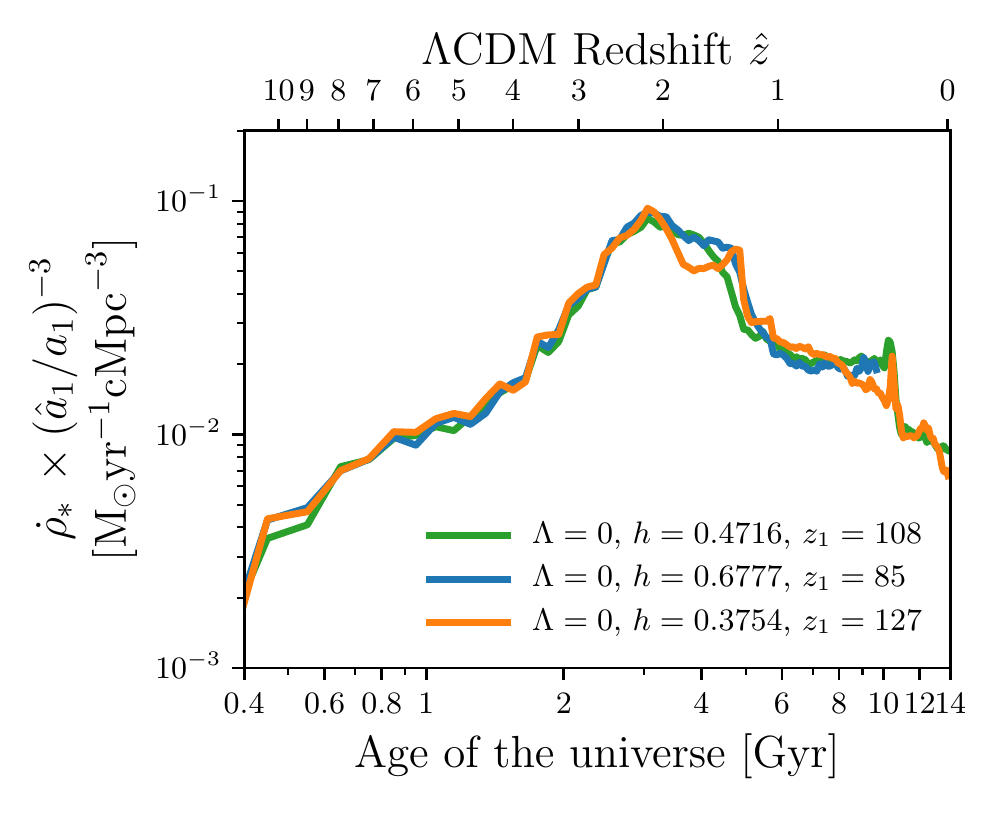}\vspace{-1.5em}
 \caption{Global SFR density for three EdS models scaled by the ratio of the initial scale factors for each model. The initial conditions for each model have been re-scaled such that the time at which we start the simulations remains unchanged, i.e. $t(a_1)=11.98 \,\mathrm{Myr}$.}
 \label{fig:Scaled_madau_all}
\end{figure}

In order to demonstrate that this re-scaling strategy works correctly, \cref{fig:Scaled_madau_all} shows the global SFR density for the three small box EdS simulations used for convergence (see \cref{tab:a}). They each represent the same physical scenario, but choose a different proper time to be ``today'', $t_0$. This has the effect of altering the values of the Hubble parameter $h$ and the redshift of the initial conditions, so that the simulations begin at proper time $t(a_1)=11.98 \,\mathrm{Myr}$ in all models. Despite the small size of the simulation boxes (hence the noisy curves), the figure shows consistent SFRs as a function of cosmic time for the three models. Therefore, our re-scaling strategy allows us to simulate any cosmological model, regardless of the value of $h$.

\begin{figure*}
 \includegraphics[width=0.95\textwidth]{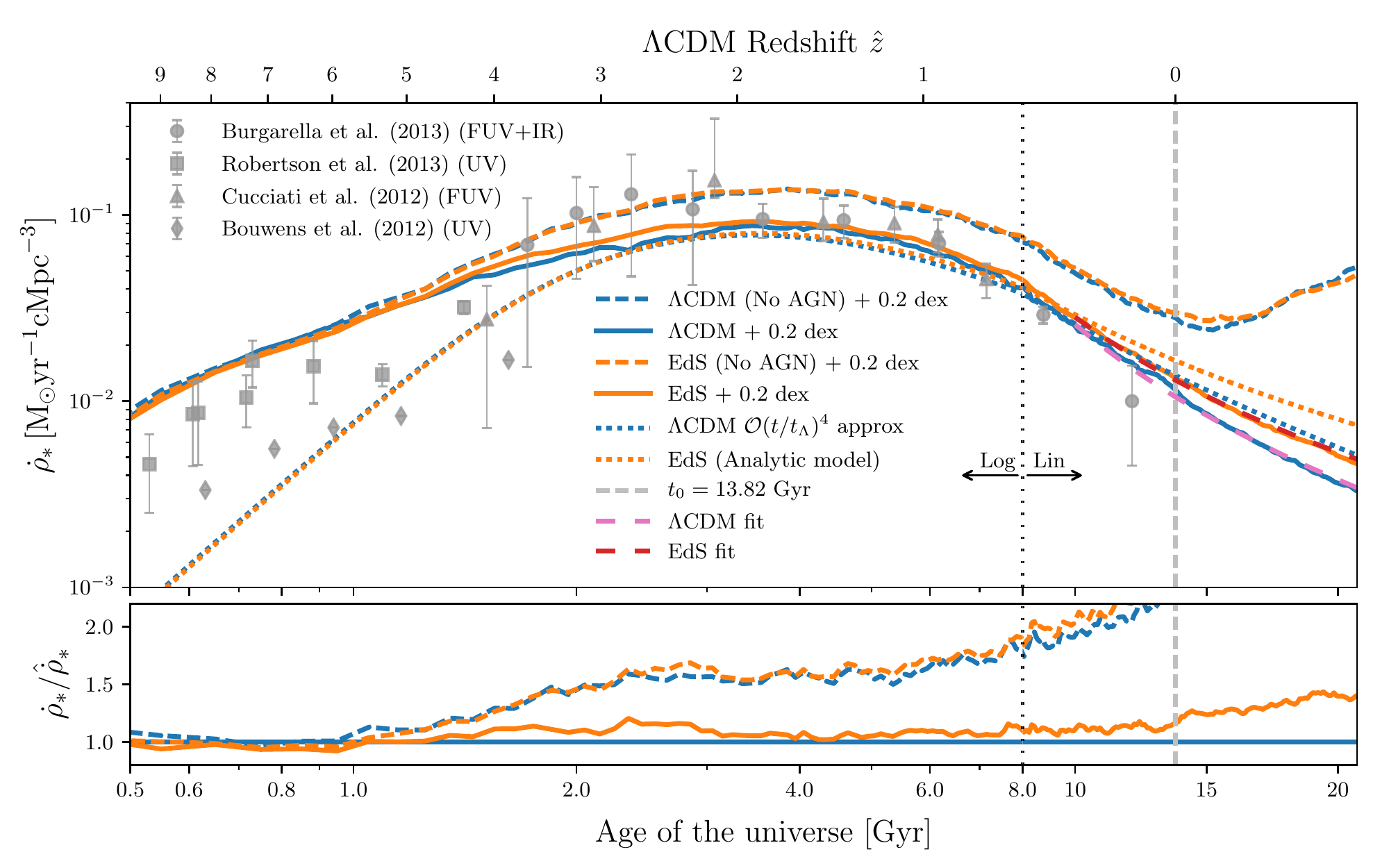}
 \caption{Global SFR densities. The \textsc{eagle} reference and the EdS models are shown in solid blue and orange lines respectively. Observational data from \protect\cite{Cucciati:2012} [FUV], \protect\cite{Bouwens:2012} [UV], \protect\cite{Robertson:2013} [UV] and \protect\cite{Burgarella:2013} [FUV + FIR] are shown as symbols. The \textsc{eagle} reference simulation (solid blue) reproduces the shape of the observed SFR density remarkably well, with a small offset of 0.2 dex at $t\gtrsim 2$ Gyr (for clarity, all \textsc{eagle} models have been shifted by 0.2 dex). The analytical model of \cref{eq:ssfr_rate_density_dm,eq:rho_int} is shown with dotted lines for both cosmologies. Power law fittings for the SFR density for $t>8$ Gyr, as per \cref{eq:fit}, for the $\Lambda$CDM and EdS models are shown in pink and red lines respectively. The horizontal time axis has been plotted in a logarithmic scale for $t \leq 8$ Gyr changing to a linear scale for $t > 8$ Gyr. The black vertical dotted line shows the transition from logarithmic to linear scale.}
 \label{fig:Scaled_madau}
\end{figure*}

\section{Results: The Evolution of Star Formation}\label{sec:Results}

\subsection{The past history of the cosmic star formation rate}\label{sec:SFR_density}

\Cref{fig:Scaled_madau} shows the global SFR density as a function of cosmic time for our simulation models. For comparison, observations from \cite{Cucciati:2012} [FUV], \cite{Bouwens:2012} [UV], \cite{Robertson:2013} [UV] and \cite{Burgarella:2013} [FUV + FIR] are shown as well. Solid lines in the figure show the evolution of the (co-moving) cosmic SFR density for the reference $\Lambda$CDM \textsc{eagle} run (blue), and for an EdS universe (orange). Dashed lines show simulation models without feedback from AGN. Dotted lines show the prediction for the cosmic SFR density using \cref{eq:ssfr_rate_density_dm}. We focus first on the evolution of the models up to the present age of the universe, $t = t_0 = 13.8$ Gyr.

In linear time, the SFR rises very rapidly and most of the plot is dominated by the slow decline (for an example, see \citealt{Furlong:2015}). Hence, in order to emphasis the growth and decline of the SFR, and to reproduce the familiar shape of the star formation history \citep{Madau:2014}, the horizontal time axis has been plotted in a logarithmic scale for $t \leq 8$ Gyr. In order to explore the SFR in detail at the present epoch and into the future, the horizontal time axis changes to a linear scale for $t > 8$ Gyr. The black vertical dotted line shows the transition from logarithmic to linear scale. For reference, the  redshifts, $\hat{z}$, for an observer at $t_0$ in the $\Lambda$CDM universe, are given along the top axis. As discussed in detail in \cite{Furlong:2015}, the reference simulation (solid blue line) reproduces the shape of the observed SFR density remarkably well, with a small offset of 0.2 dex at $t\gtrsim 2$ Gyr. While the simulations agree reasonably well with the observational data at redshifts above 3, we caution that these measurements are more uncertain.

Remarkably, the shape of the cosmic SFR history is very similar for both the $\Lambda$CDM and EdS models: the SFR density peaks ${\approx}3.5$ Gyr after the Big Bang and declines slowly thereafter. The similarity of the universes prior to the peak is expected, since the $\Lambda$ term in the Friedman equation is sub-dominant in both cases. At later times, however, we might naively have expected the decline to be more pronounced in the $\Lambda$CDM cosmology, since the growing importance of the $\Lambda$ term slows the growth of density perturbations. 

From \cref{fig:D_t}, the linear growth factors of the two cosmological models differ by ${\approx}10\%$ at the present time, and so we might have expected a similar difference in the (co-moving) cosmic SFR density (${\approx} 15 \%$, read from \cref{fig:Scaled_madau}). This naive expectation is not borne out because of the complexity of the baryonic physics. Because of stellar and AGN feedback, haloes have an ample reservoir of cooling gas that is able to power further star formation regardless of the change in the cosmic halo growth rate. 

Our simulation demonstrates that the existence of $\Lambda$ does play a small role in determining the (co-moving) cosmic SFR density. However, these differences are minor. In order to put the differences into context, we compare with a pair of simulations in which the BH feedback is absent. These runs are shown as dashed lines in the plot. We focus here on the behaviour before $t=13.8$ yrs. As can be seen, the absence of AGN feedback has a dramatic effect on the shape of the cosmic star formation density \citep{Schaye_et_al:2010, vandeVoort:2011}. Interestingly, however, while the normalisation of the SFR density is considerably higher, the time of the peak is similar. BH feedback is not solely responsible for the decline in star formation after $ t \approx 3.5$ Gyr. This hints that the existence of the peak results from the interaction of the slowing growth rates of haloes (after the peak) and the star formation timescale (set by the ISM physics) which limits the rate at which the galaxy can respond to convert in-falling material into stars (before the peak).

The SFR history predicted by the simple model developed in \cref{sec:approx} (dotted curves) is in remarkable agreement with the observational data and predicts the relative difference of the cosmological simulations, both at the present time, and into the future. We want to emphasise that the model is not a parametric fit to the data, but rather an analytical model derived from a simple relation of star formation to halo mass accretion. 

\subsection{The future of the cosmic star formation rate ($t > 13.8$ Gyr)}

In order to explore whether the relative SFR densities will diverge as the impact of $\Lambda$ becomes more pronounced, we ran the simulations for both cosmological models into the future (i.e. beyond a cosmic time of $t = 13.8$ Gyr). As the simulations run into the future, the small differences seen at $t = 13.8$ Gyr become larger, reading an ${\approx}40\%$ difference at $t = 1.5\times t_0 = 20.7$ Gyr, which is in agreement with the predictions shown in \cref{fig:madau_prediction}.

The decline in the SFR density can be approximated by a power law for both the reference $\Lambda$CDM and EdS models (red and pink dashed lines),
\begin{equation}\label{eq:fit}
	\dot{\rho}_* = a(ct)^{-k},
\end{equation}
with the parameters $a$, $c$ and $k$ given in \cref{tab:fit}. We used the reduced chi-squared statistic for goodness of fit testing. We use this fitting function to extrapolate the results from the simulations further into the future.

\begin{table}
\centering
\caption{Power law parameter fitting for the median SFR shown in \cref{fig:Scaled_madau}.}\label{tab:fit}
\begin{tabular}{lccc}
Model        & a    & c    & k    \\ 
& $[\mathrm{M}_\odot \mathrm{yr}^{-1} \mathrm{cMpc}^{-3}]$ & $[\mathrm{Gyr}^{-1}]$ & \\ \hline 
$\Lambda$CDM & 5.42 & 0.81 & 2.77 \\ 
EdS          & 3.99 & 0.94 & 2.41 \\ \hline 
\end{tabular}
\end{table}

We note here a striking feature of the universes with no BH feedback: the SFR increases again in the future, both for the $\Lambda$CDM and EdS cosmologies. In \cref{sec:gal_prop} we discuss how this effect originates from massive galaxies ($M_* > 10^{11} \mathrm{M}_\odot$) that rejuvenate in the future. As it is not clear from the simulation whether the SFR will continue to rise, or when it would start declining, we have not fitted any functional form to the ``No AGN'' models.

\begin{figure}
 \includegraphics[width=0.48\textwidth]{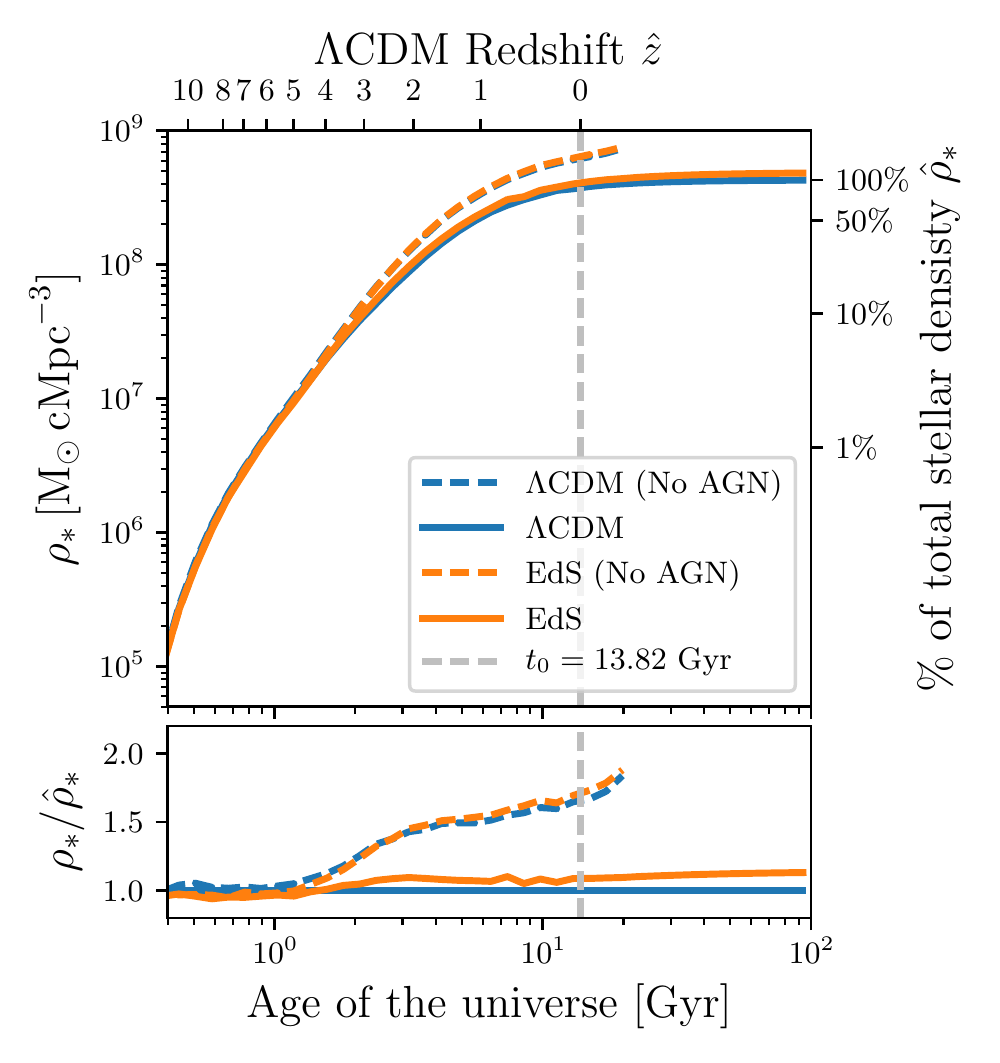}
 \caption{The stellar mass density as a function of time in the \textsc{eagle} simulation models. The colour coding is the same as in \cref{fig:Scaled_madau}. The right-hand axis represents the percentage of the total stellar density compared to the $\Lambda$CDM model. Both the $\Lambda$CDM and EdS models have already produced most of the stars in the universe by the present day, building up very little stellar mass into the future. The models without AGN feedback quickly deviate from the reference model, producing almost twice the mass in stars by the end of the simulation (${\approx}20.7$ Gyr). The figure shows that the effect of dark energy on the overall star formation is negligible.}
 \label{fig:Stellar_densiy}
\end{figure}

\begin{figure*}\centering
 \includegraphics[width=0.95\textwidth]{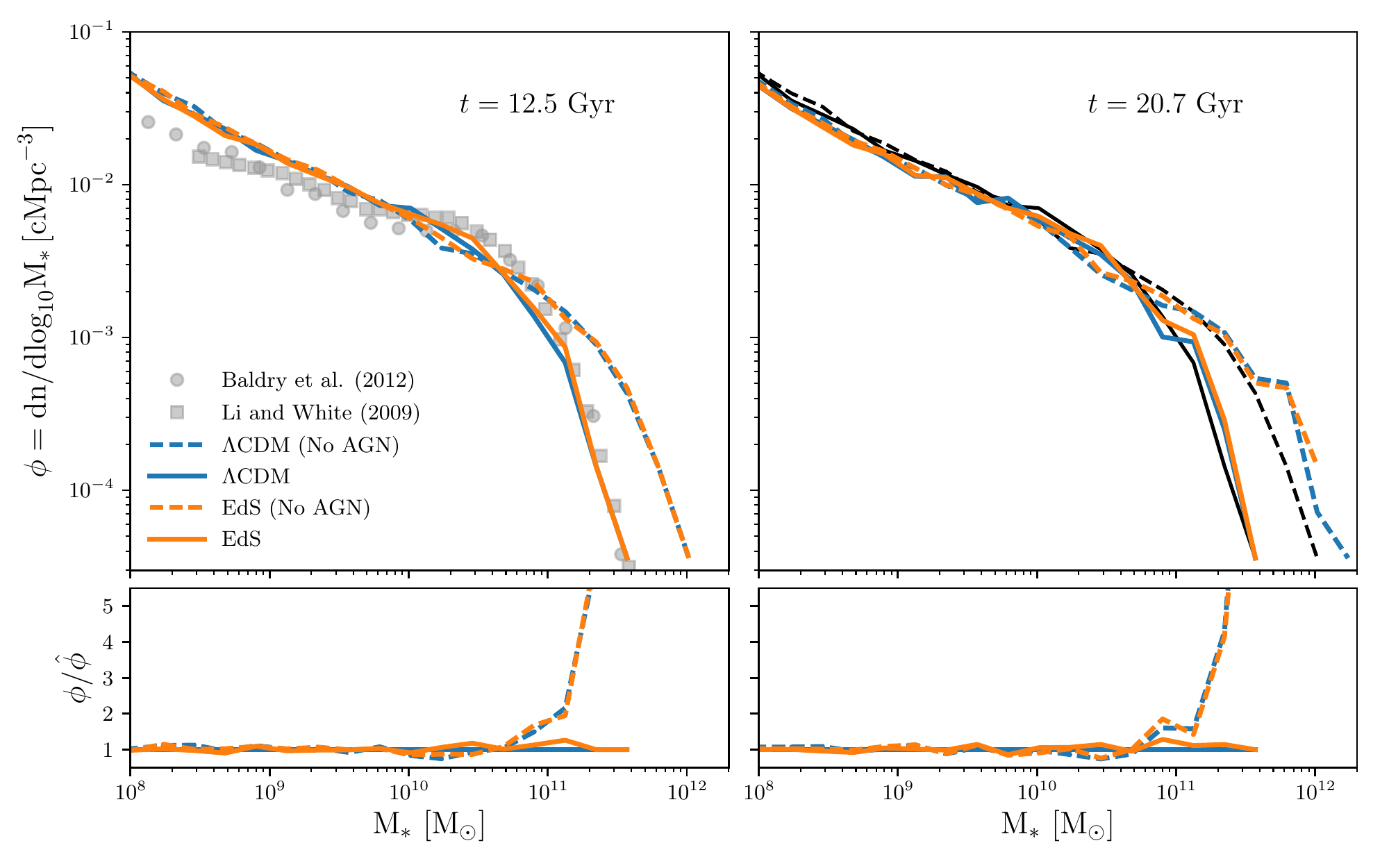}
 \vspace{-1.5em}
 \caption{The GSMF at $t = 12.5$ Gyr, equivalent to redshift $\hat{z}=0.1$ for an observer at the present time in a $\Lambda$CDM universe (left), and $t=1.5 \times t_0$ (right) for the \textsc{eagle} simulation models. The colour coding is the same as in \cref{fig:Scaled_madau}. Observational data from \protect\cite{Baldry:2012} and \protect\cite{LiWhite:2009} is shown as symbols. The reference and ``No AGN'' $\Lambda$CDM models at $t = 12.5$ Gyr are plotted in the right panel for reference (solid and dashed black lines respectively). The effect of dark energy on the GSMF is negligible, with very little evolution into the future. As expected, the models without AGN feedback predict a higher number density of massive galaxies ($M_* > 10^{11} \mathrm{M}_\odot$). This effect becomes more significant into the future, going from ${\approx}0.7$ dex to ${\approx}1.1$ dex.}
 \label{fig:SMF_0}
\end{figure*}

\begin{figure*}\centering
 \includegraphics[width=0.95\textwidth]{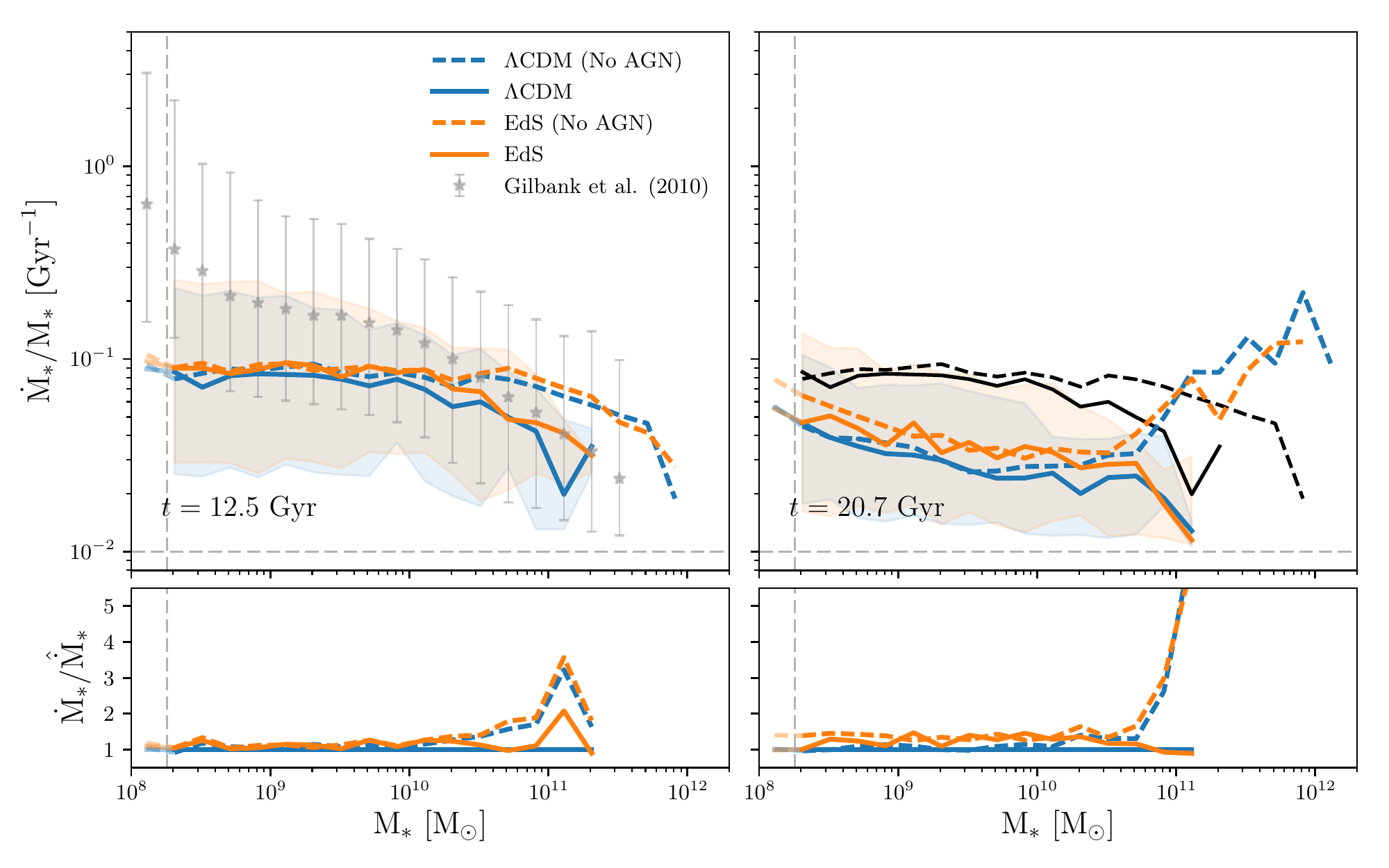}
 \vspace{-1.5em}
 \caption{The SSFR of star forming galaxies at $t = 12.5$ Gyr, equivalent to redshift $\hat{z}=0.1$ for an observer at the present time in a $\Lambda$CDM universe (left), and $t=1.5 \times t_0$ (right) for the \textsc{eagle} simulation models. The colour coding is the same as in \cref{fig:Scaled_madau}. Observational data from \protect\cite{Gilbank:2010} is shown as symbols. The reference and ``No AGN'' $\Lambda$CDM models at $t = 12.5$ Gyr are plotted in the right panel for reference (solid and dashed black lines respectively). The solid curves show the median relation for star forming galaxies, defined as those with an SSFR above the limit specified by the horizontal dashed line ($10^{-2} \mathrm{Gyr}^{-1}$). The faint shaded regions enclose the 10th to 90th percentiles for the $\Lambda$CDM and EdS Models. Lines are light coloured when the stellar mass falls below that corresponding to 100 baryonic particles, to indicate that resolution effects will be important. The figure shows that the effect of dark energy on the GSMF is negligible. The models without AGN feedback predict a higher SSFR for massive galaxies ($M_* > 10^{10} \mathrm{M}_\odot$). The right panel shows that the overall SSFR drops from $t = 12.5$ Gyr to $t=1.5 \times t_0$. For the ``No AGN'' models, however, the SSFR increases for massive galaxies ($M_* > 10^{11} \mathrm{M}_\odot$).}
 \label{fig:SSFR_0}
\end{figure*}

\subsection{The stellar mass density}

To study the build up of stellar mass, we present the growth in stellar mass density, $\rho_*$, across cosmic time in \cref{fig:Stellar_densiy}. The colour coding is the same as in \cref{fig:Scaled_madau}. The lower panel shows the ratio of the stellar density compared to the reference $\Lambda$CDM model. \cite{Furlong:2015} show that the reference \textsc{eagle} simulation is in good agreement with the observed growth of stellar mass across cosmic time.  In contrast to \cite{Furlong:2015}, where the stellar mass density was obtained from aperture measurements to facilitate comparison with observations, we calculate $\rho_*$ by integrating the SFR density from \cref{fig:Scaled_madau},
\begin{equation}\label{eq:stellar_density}
	\rho_* = \int_0^t \dot{\rho}_* \dd t^\prime,
\end{equation}
in order to provide an estimate of the total mass of stars produced by the universe\footnote{As we are interested in total mass of stars \emph{produced} by the universe, \cref{eq:stellar_density} ignores stellar mass loss.}. For the models with AGN feedback, we have extrapolated the power law fit described in \cref{sec:SFR_density} far into the future, up to 10 trillion years, and considered this as the ``Total stellar mass density'' of the universe. As suggested by the analytic model in \cref{eq:ssfr_rate_density_dm}, in universes with feedback from star formation and AGN, the cosmic SFR density is expected to continue decreasing into the future. At late times, the SFR becomes orders of magnitude lower than that of the peak at $3.5$ Gyr. \Cref{fig:Stellar_densiy} shows that the total stellar mass density is dominated by the contribution from the peak in star formation and reaches a plateau at $t{\approx}t_0$. Hence, the formal uncertainties in the extrapolation into the future are unimportant for the predicted total stellar mass density. The right-hand axis of \cref{fig:Stellar_densiy} represents the percentage of the total stellar density compared to $\Lambda$CDM. For the reference $\Lambda$CDM model, the universe has already produced most (${\approx}88\%$) of its eventual stellar mass by the present day, adding up very little stellar mass into the future. Although there is no $\Lambda$ to slow down the formation of cosmic structure, the EdS model closely resembles a $\Lambda$CDM cosmology with only ${\approx}15\%$ more stellar mass produced. 

As discussed in \cref{sec:future}, a universe with $\Lambda$ has a very different expansion history  compared to one without dark energy. This produces a different growth of density perturbations, in particular into the future (see \cref{fig:a_t,fig:D_t}). Nevertheless, as seen in \cref{fig:Stellar_densiy}, since both cosmologies have already produced most of the stars in the universe by the present day, when the contribution of $\Lambda$ is becoming increasingly important, the effect of dark energy on the overall star formation, is negligible. 

In contrast, the models without feedback from BHs (dashed curves) quickly deviate from the reference model, starting from $t \sim 1$ Gyr, producing almost twice the mass in stars by the end of the simulations, $20.7$ Gyr after the Big Bang.

\subsection{Other galaxy population properties}\label{sec:gal_prop}
In this section we will compare the galaxy population properties of the two simulation models at the present time, and into the future. In particular, we compare the galaxy stellar mass function (GSMF), and the specific star formation rate (SSFR) of galaxies. To compare with observational data, in each of the figures discussed below, the left panel shows properties at $t = 12.5$ Gyr, equivalent to redshift $\hat{z}=0.1$ for an observer at the present time in a $\Lambda$CDM universe. The right panel shows the same property but at $t=1.5 \times t_0 = 20.7$ Gyr. To guide the eye, each property for the reference $\Lambda$CDM model at $t = 12.5$ Gyr is plotted with a black line in each plot at $t=1.5 \times t_0 = 20.7$ Gyr. Finally, we have also included the ratios of each quantity to the reference $\Lambda$CDM model at the corresponding time at the bottom of each panel.

\subsubsection{The galaxy stellar mass function}

The effect of $\Lambda$ on the GSMF can be seen in \cref{fig:SMF_0}. The colour coding is the same as in \cref{fig:Scaled_madau}. For comparison, observational data from \cite{Baldry:2012} and \cite{LiWhite:2009} is shown as well. As discussed in \cite{Schaye:2015}, the observed GSMF at redshift 0.1 was used to infer the free parameters of the subgrid physics used in the simulation. The reference \textsc{eagle} model reproduces  the shape of the observed GSMF reasonably well, with a slight under-abundance of galaxies at its knee. 

\Cref{fig:SMF_0} shows that the effect of $\Lambda$ on the GSMF is negligible, with very little evolution into the future. The models without AGN feedback predict a higher number density of massive galaxies ($M_* > 10^{11} \mathrm{M}_\odot$) compared to the models with AGN feedback. This effect becomes more significant into the future, going from 0.7 dex to 1.5 dex. The origin of this difference is explored in the next section. 

\subsubsection{Specific star formation rates}

Galaxies can be broadly classified into largely distinct star-forming and passive populations according to their SSFR,
\begin{equation}
	\mathrm{SSFR} = \frac{\dot{M}_*}{M_*}.
\end{equation}

For star-forming galaxies, there is a well-defined star forming sequence, with SSFR observed to be approximately constant as a function of stellar mass \citep[e.g.][]{Noeske:2007,Karim:2011}. \Cref{fig:SSFR_0} shows the SSFR for star-forming galaxies in the simulations as a function of galaxy stellar mass at the present day, and into the future. The colour coding is the same as in \cref{fig:Scaled_madau}. The horizontal dotted lines correspond to the SSFR cut ($10^{-2}$ Gyr$^{-1}$) used to separate star forming from passive galaxies in our Universe. For comparison, observational data from \cite{Gilbank:2010} is show as well. \cite{Furlong:2015} show that the SFFR in the reference simulations at the present day is similar to observations in the local universe, with an offset of 0.3 dex. This is possibly consistent with the systematic uncertainties in the calibration of the  observation diagnostics. At low masses there is an increase in SSFR with stellar mass; however, this has been found to be a resolution-dependent effect. Hence, we have plotted the results with lighter coloured lines \citep[similar to][]{Schaye:2015}. The models without feedback from AGN have higher SSFR for $M_* > 10^{10} \mathrm{M}_\odot$, whereas the effect of $\Lambda$ on the SSFR of galaxies is negligible.

\Cref{fig:SSFR_0} shows the galaxy population property that has the strongest evolution into the future. We find that over the next 6.8 Gyr the SSFR will drop by ${\approx}0.4$ dex. 

Interestingly, the models without AGN feedback predict an increase of the SSFR of galaxies with $M_* > 10^{11} \mathrm{M}_\odot$ in the future. The figure shows that the increase in SFR shown in \cref{fig:Scaled_madau}, and the higher number density of massive galaxies in \cref{fig:SMF_0}, originate from massive galaxies that rejuvenate in the future. A plausible explanation for this phenomenon is that, according to simple radiative cooling models, in massive galaxies and clusters, a hot gaseous atmosphere should lose energy by the emission of radiation, and if there is no heating mechanism to compensate the cooling (e.g. AGN feedback), cooling flows should form \citep{Fabian:1994,Peterson:2006}, triggering star formation. This result will be explored in more detail in a follow-up paper.  

\section{Discussion and Conclusions}\label{sec:con}

In this paper, we explored the dependence of the star formation history of the universe on the existence of a cosmological constant and feedback from accreting BHs. We base our results on the \textsc{eagle} simulation code, that has been shown to compare favourably to observational data, and thus, to provide a good description of the formation of galaxies in our Universe. Feedback from supermassive BHs has been shown to be a key ingredient in achieving this match by suppressing star formation in massive haloes \citep[e.g.][]{Bower:2006,Harrison:2017}, while the accelerating expansion rate of the Universe suppresses the accretion rates of haloes at late times \citep[e.g.][]{Jenkins:1998,Huterer:2015}. Our study allows us to assess the relative importance of these ingredients.

The universes that we consider are indistinguishable at early times. They share a common epoch of equality and recombination, and have equal amplitudes and spectrum of density fluctuations at early times. We take care to compare the evolution of models with equivalent starting points, and to demonstrate that the simulation code correctly scales the different values of the present-day expansion rate (Hubble parameter). When comparing the universes, it is important that we compare properties at a fixed cosmic time. Since the processes of stellar (and biological) evolution provide a common clock, independent of the large-scale cosmological expansion, these provide an astrophysically relevant comparison. 

We have also developed an analytic model derived from a simple relation of star formation to halo mass accretion rate. Despite its simplicity, the model reproduces the overall shape of evolution of the cosmic SFR density. The model and the simulations allow us to explore the effect of a cosmological constant term on the cosmic SFR density.

Our main conclusions are as follows:

\begin{itemize}
	\item We find that the existence of the cosmological constant has little impact on the star formation history of the Universe. The SFR is suppressed by ${\approx}15\%$ at the present time, and we find that the properties of galaxies are almost indistinguishable in the two universes.
	\item To explore whether this is due to the relatively recent dominance of the dark energy density in our Universe, we continued the simulations $6.8$ Gyr into the future. Even after this time, the co-moving SFR densities differ only by ${\approx}40\%$. Clearly, the cosmological constant has only a marginal effect on the stellar content of the Universe.
	\item Using the analytic model, we can recognise that the existence of the peak in the SFR density results from the interaction of the star formation efficiency (set by the ISM physics) which limits the rate at which the galaxy can respond to convert in-falling material into stars, the relative abundance of efficiently star forming haloes (i.e. of masses $\approx 10^{12} \Msol$), and only at late times, the slowing growth rates of haloes due to the cosmological constant.
	\item By extrapolating fits to the evolution of the co-moving SFR density into the future, we show that, in our Universe, more than ${\approx}88\%$ of the stars that will ever be produced, have already been formed by the present cosmic time. In the absence of dark energy, only ${\approx}15\%$ more stellar mass would have been formed in the same time. The difference is small, bringing into question whether the `coincidence problem' (the comparable energy densities of matter and dark energy) can be explained by an anthropic argument: the existence of dark energy (at the observed value) has negligible impact on the existence of observers or the ability of humanity to observe the cosmos. In \cite{Barnes:2018} we explore this argument in more detail by considering a wider range of $\Lambda$ values, and determining the likelihood distribution of possible $\Lambda$ values conditioning the existence of observers.
	\item In comparison, the existence of BHs has a major impact on the Universe. In the absence of AGN feedback, the co-moving SFR density is enhanced by a factor of 2.5 at the present day.
	\item Even in a universe without BHs or dark energy, we find that the co-moving SFR density peaks at 3.5 Gyr ($z{\approx}2$ according to a present-day observer in our Universe). The decline in star formation is however slower at more recent times.
	\item For hypothetical universes without feedback from accreting BHs, there is a \emph{comeback} of SFR, which increases again in the future. This effect originates from massive galaxies ($M_* > 10^{11} \mathrm{M}_\odot$) that rejuvenate as there is no heating mechanism to compensate the cooling, in turn, triggering star formation.
\end{itemize}

\section*{Acknowledgements}

We are grateful to all members of the Virgo Consortium and the \textsc{eagle} collaboration who have contributed to the development of the codes and simulations used here, as well as to the people who helped with the analysis. We thank Alejandro Benitez-Llambay for the visualisation software \textsc{py-sphviewer} \citep{sphviewer}.

This work was supported by the Science and Technology Facilities Council (grant number ST/P000541/1); European Research Council (grant number GA 267291 \enquote{Cosmiway}) and by the Netherlands Organisation for Scientific Research (NWO, VICS grant 639.043.409). RAC is a Royal Society University Research Fellow.

This work used the DiRAC Data Centric system at Durham University, operated by the Institute for Computational Cosmology on behalf of the STFC DiRAC HPC Facility (\url{http://www.dirac.ac.uk}). This equipment was funded by BIS National E-infrastructure capital grant ST/K00042X/1, STFC capital grant ST/H008519/1, and STFC DiRAC Operations grant ST/K003267/1 and Durham University. DiRAC is part of the National E-Infrastructure. We acknowledge PRACE for awarding us access to the Curie machine based in France at TGCC, CEA, Bruy\`{e}res-le-Ch\^{a}tel. 

Jaime Salcido gratefully acknowledges the financial support from the Mexican Council for Science and Technology (CONACyT), fellow no. 218259.




\bibliographystyle{mnras}
\bibliography{biblio_new} 




\appendix
\renewcommand{\thetable}{T\arabic{table}}
\renewcommand{\thefigure}{\thesection\arabic{figure}} 


\bsp	
\label{lastpage}
\end{document}